\documentclass[fleqn,usenatbib]{mnras}

\usepackage{newtxtext,newtxmath}

\usepackage[T1]{fontenc}

\DeclareRobustCommand{\VAN}[3]{#2}
\let\VANthebibliography\thebibliography
\def\thebibliography{\DeclareRobustCommand{\VAN}[3]{##3}\VANthebibliography}

\usepackage{graphicx}	
\usepackage{amsmath}	

\usepackage{graphicx}	
\usepackage{amsmath}	

\usepackage{listings}
\usepackage{xcolor}
\usepackage{siunitx}
\usepackage{xspace}
\usepackage[normalem]{ulem}	
\usepackage{array}	

\newcommand{\synthesizer}{\textsc{synthesizer}\xspace}

\newcommand{\unyt}{\textsc{unyt}\xspace}
\newcommand{\emodel}{\texttt{EmissionModel}\xspace}

\newcommand{\grid}{\texttt{Grid}\xspace}
\newcommand{\grids}{\texttt{Grids}\xspace}

\newcommand{\app}[1]{Appendix~\ref{sec:#1}}

\newcommand{\fig}[1]{Figure~\ref{fig:#1}}
\renewcommand{\sec}[1]{Section~\ref{sec:#1}}

\newcommand{\Eq}[1]{Equation~\ref{eq:#1}}
\newcommand{\Fig}[1]{Figure~\ref{fig:#1}}

\newcommand{\Tab}[1]{Table~\ref{tab:#1}}

\DeclareSIUnit\angstrom{\text {Å}}

\title[How bright can a galaxy be?]{How bright can a galaxy be? Extreme value predictions for high-redshift survey observables}
\author[C. C. Lovell et al.]{Christopher C. Lovell,$^{1,2}$\thanks{E-mail: chris.lovell.astro@gmail.com (CCL)}
Theodore Martin,$^{3}$
Sandro Tacchella,$^{1,2}$
Christian K. Jespersen$^{4,5,6}$ and
\newauthor
Ian Harrison$^{7}$
\\
$^{1}$Kavli Institute for Cosmology, University of Cambridge, Madingley Road, Cambridge CB3 0HA, UK\\
$^{2}$Institute of Astronomy, University of Cambridge, Madingley Road, Cambridge CB3 0HA, UK\\
$^{3}$Institute of Cosmology and Gravitation, University of Portsmouth, Burnaby Road, Portsmouth PO1 3FX, UK\\
$^{4}$David A. Dunlap Department of Astronomy \& Astrophysics, University of Toronto, Toronto, ON M5S 3H4, Canada\\
$^{5}$Department of Statistical Sciences, University of Toronto, Toronto, ON M5S 3G3, Canada\\
$^{6}$Canadian Institute for Theoretical Astrophysics, University of Toronto, Toronto, M5S 3H8, Ontario, Canada\\
$^{7}$Department of Physics, School of Physical, Chemical and Environmental Sciences, Cardiff University, Cardiff CF24 3AA, UK
}

\date{Accepted XXX. Received YYY; in original form ZZZ}

\pubyear{2026}

\begin{document}
\label{firstpage}
\pagerange{\pageref{firstpage}--\pageref{lastpage}}
\maketitle

\begin{abstract}
Assessing the significance of extreme galaxy candidates has typically relied on redshifts and stellar / halo masses inferred through inverse modelling, which carries substantial systematic uncertainty.
We combine the halo mass function with a conditional model for galaxy emission and apply the Extreme Value Statistics (EVS) formalism to predict the distribution of the observed flux of the brightest galaxy expected in a given survey area, redshift interval and photometric band.
We find the choice of parametric star formation history (SFH) dominates the modelling uncertainty, introducing a spread approaching $\sim 1 \, \mathrm{dex}$ by $z \sim 18$, whilst the stellar population synthesis model, initial mass function and metallicity each contribute $\lesssim 0.3 \, \mathrm{dex}$.
The luminous $z \gtrsim 9$ candidates in COSMOS-Web cluster about the median prediction for the area they were drawn from, and lie at least an order of magnitude below the $3\sigma$ limit, a margin comparable to the SFH systematic.
Of the observational samples reaching beyond $z \sim 13$, the two spectroscopically confirmed galaxies at $z \simeq 14$ lie close to the median prediction for their area, whereas photometric candidates at $z \gtrsim 16$ reach $+4\sigma$ (a probability to exceed of $\sim 10^{-5}$), the brightest lying at the limit obtained by converting all halo baryons into stars.
Since the prediction bounds the flux far more tightly than the photometry bounds the redshift, such offsets are most naturally resolved by errors in the inferred redshifts.
This approach allows extreme candidates to be assessed directly in the observational plane, making explicit the modelling assumptions common to forward and inverse approaches.
\end{abstract}

\begin{keywords}
galaxies: high-redshift -- galaxies: luminosity function, mass function -- galaxies: statistics -- galaxies: formation -- methods: statistical
\end{keywords}



\section{Introduction}
\label{sec:intro}

JWST has revealed a range of unexpected and exciting galaxy populations, including a number of extremely bright galaxy candidates at very high redshifts \citep{finkelsteinLongTimeAgo2022,naiduTwoRemarkablyLuminous2022a,castellanoEarlyResultsGLASSJWST2023,castellanoPushingJWSTExtremes2025a,labbePopulationRedCandidate2023,caseyCOSMOSWebIntrinsicallyLuminous2024,robertsonEarliestGalaxiesJADES2024,perez-gonzalezRiseGalacticEmpire2025a,mcleodSearchFirstGalaxies2026a}. 
The abundance of luminous sources at $z \gtrsim 10$ exceeds the extrapolation of pre-JWST ultraviolet luminosity functions, and the bright end evolves more slowly with redshift than had been anticipated \citep{donnanEvolutionGalaxyUV2023,harikaneComprehensiveStudyGalaxies2023,wilkinsFirstLightReionization2023a,chworowskyEvidenceShallowEvolution2024,whitler9GalaxyUV2025}.
Inverse modelling through SED fitting initially suggested that many of these sources had redshifts and stellar masses that were in tension with our models of astrophysics and, in the most extreme cases, the concordance cosmological model itself \citep{lovellExtremeValueStatistics2023,boylan-kolchinStressTesting$Lambda$CDM2023}.

Since then, updates to the data calibration and improvements to the modelling procedure have reduced this tension somewhat \citep{masonBrightestGalaxiesCosmic2023,yungAreUltrahighredshiftGalaxies2024}.
Spectroscopic follow-up has shown a number of the most extreme photometric candidates to be lower-redshift dusty or evolved interlopers, brown dwarfs, or AGN/LRD candidates \citep{zavalaDustyStarburstsMasquerading2023,arrabalharoConfirmationRefutationVery2023,hainlineJWSTAdvancedDeep2026a,bradacTwoExcitingHighredshift2026}, and re-analyses including longer-wavelength constraints typically recovered lower stellar masses than were first reported \citep{desprezLCDMNotDead2024,wangTrueNumberDensity2024,lapasiaStellarMassesOptically2026a}.
A number of astrophysical explanations have also been proposed that accommodate the observed abundances within $\Lambda$CDM, including feedback-free or otherwise highly efficient starbursts \citep{dekelEfficientFormationMassive2023,liFeedbackfreeStarburstsCosmic2024, somervilleDensityModulatedStar2025}, bursty star formation and the associated stochastic variability in the UV luminosity \citep{sunBurstyStarFormation2023,shenImpactUVVariability2023a,gelliImpactMassdependentStochasticity2024}, reduced dust attenuation in early, compact systems \citep{ferraraSuperearlyJWSTGalaxies2024}, and a contribution to the continuum from accreting black holes \citep{hegdeHiddenPopulationActive2024a}.
Whether any tension remains is therefore, to a large degree, a question about the chain of modelling assumptions connecting an observed flux to a dark matter halo mass, and any uncertainties on the redshift estimate.

Extreme Value Statistics (EVS) provides a natural framework in which to pose this question.
Rather than asking whether the abundance of an entire population is consistent with a model, EVS gives the full probability distribution of the single most extreme object expected in a given sample \citep{gumbelStatisticsExtremes1958,katzExtremeValueDistributions2000}.
This makes it well suited to assessing individual record-breaking objects, since a single object detection in excess of the prediction is, in principle, falsifying.
The approach has been applied extensively to the most massive galaxy clusters \citep{davisMostMassiveHaloes2011,harrisonExactExtremeValue2011,harrisonTestingCosmologyExtreme2012,holzMostMassiveObjects2012,waizmannApplicationExtremeValue2012}, and more recently to the most massive and most luminous galaxies at high redshift \citep{lovellExtremeValueStatistics2023,heatherExtremevalueModellingBrightest2024a,enriquez-vargasEmpiricalEstimatesHow2026}.

In \cite{lovellExtremeValueStatistics2023} we made predictions for the most massive galaxy expected in a given survey, using the EVS formalism.
By assuming a form for the halo mass function, a cosmic baryon fraction and stellar fraction we placed both upper and lower limits on the most extreme mass halo and galaxy.
\cite{jespersenSignificanceRareObjects2025} showed the impact of cosmic variance on these predictions from pencil beam / small area surveys, showing that for a single JWST pointing the median predicted mass can drop by over 1 dex at $z > 5$.
The same clustering that drives this variance also implies that the most extreme sources are preferentially found in overdense, protocluster-like environments \citep[\textit{e.g.}][]{castellanoEarlyResultsGLASSJWST2023}, so that conditioning the prediction on the observed density of the field in which a candidate is found can substantially modify its inferred significance \citep{jespersenExplainingUltramassiveQuiescent2025}.
Other studies extended the analysis to galaxies with extreme \textit{assembly} histories, by tracking their inferred star formation histories back to ask whether there was any tension in their masses formed at some earlier time \citep{carnallJWSTEXCELSSurvey2024}; these approaches require additional assumptions to be made about the merger history, to account for hierarchical assembly \citep{cochraneHierarchicalAssemblyImpedes2025a,turnerAgedatingEarlyQuiescent2025}.

However, all of these studies use stellar and halo masses inferred through inverse modelling, typically through spectral energy distribution (SED) fitting of the multi-wavelength electromagnetic emission of individual galaxies from photometric and, in more limited cases, spectroscopic data.
Inverse modelling necessitates many assumptions that can impact the derived properties to varying degrees \citep{conroyModelingPanchromaticSpectral2013,lejaDerivingPhysicalProperties2017,iyerSpectralEnergyDistributions2026}.
These include the details of the source populations, such as the assumed initial mass function (IMF), and the stellar metallicity, as well as properties of intervening attenuating matter such as nebular gas and dust in the interstellar medium.
There are often strong degeneracies between the parameters describing these properties, leading to large uncertainties on their derived properties \citep{robothamProGenyNewSimple2025,bellstedtPROGENYIIImpact2025}.
At high redshift these issues are particularly acute: the templates and priors adopted are calibrated on lower-redshift populations \citep{steinhardtTemplatesFittingPhotometry2023}, and young, luminous stellar populations can outshine older stars, biasing the recovered stellar masses \citep{tacchellaStellarPopulationsGalaxies2022,narayananOutshiningRecentStar2024,harveySpotlightSystematicAssessment2025a}.
Different codes and parameter choices applied to the same photometry can therefore return masses differing by an order of magnitude, which directly affects any statement of tension.

An alternative to inverse modelling is \textit{forward modelling}, where the observed properties of galaxies are predicted, and comparisons between models and observations are done directly in this observational space.
Whilst many of the same choices and assumptions must be made when building forward models, the impact of these individual assumptions can arguably be explored more explicitly, and the computational intricacy of solving the inverse problem can largely be circumvented.
Comparison in the space of the observable also avoids the propagation of priors adopted during the fitting of individual sources into any population-level statement.

In this paper we extend the analysis in \cite{lovellExtremeValueStatistics2023} by forward modelling the star formation histories and emission from galaxies in order to predict the brightest galaxy, in a given photometric band in a given survey.
We use the \synthesizer{} package \citep{lovellSynthesizerSoftwarePackage2025,roperSynthesizerSyntheticObservables2026} to compute the emission, which allows us to vary the assumed star formation history, stellar population synthesis model, IMF and nebular emission independently, and quantify the contribution of each to the uncertainty in the predicted flux.
This paper is arranged as follows.
We describe the EVS formalism and its application to halo and galaxy catalogues in \sec{evs}.
Our method for modelling star formation histories, and forward modelling emission with the \synthesizer{} code is described in \sec{obs}.
We then present our predictions in \sec{results}, discuss their physical plausibility in \sec{discussion}, and summarise our conclusions in \sec{conclusions}.
Throughout we assume a $\Lambda$CDM cosmology with the \cite{planckcollaborationPlanck2015Results2016} parameters, unless otherwise stated.

\section{Extreme value predictions for halo and stellar mass}
\label{sec:evs}

The predictions in this paper are intended to test whether observed galaxies are in tension with $\Lambda$CDM and our leading model for galaxy formation.
Each relies on a number of modelling choices, which we either vary over a plausible range or set to their conservative limit.
We describe a choice as \textit{conservative} if it raises the predicted flux, so that tension is harder to claim, and as \textit{liberal} if it lowers it.

Extreme Value Statistics \citep[EVS;][]{gumbelStatisticsExtremes1958, katzExtremeValueDistributions2000} is introduced and described in detail in \cite{harrisonExactExtremeValue2011,harrisonTestingCosmologyExtreme2012,lovellExtremeValueStatistics2023}; here we provide a brief review of the approach as applied to the prediction of the most extreme halo masses and galaxy stellar masses.

\subsection{The extreme value distribution}
\label{sec:evs_formalism}

Consider a sequence of $N$ random variates $\{ M_{i} \}$ drawn from a cumulative distribution function (CDF), $F(m)$.
There will be a largest value of the sequence, $M_{\mathrm{max}} \equiv \mathrm{sup} \{ M_{1} ... M_{N}\}$.
Assuming all variables are mutually independent and identically distributed (IID), the probability all deviates are less than or equal to some value $m$ is given by:
\begin{align}
\Phi (M_{\mathrm{max}} \leqslant m;\, N) &= F_{1} (M_{1} \leqslant  m) \; ... \; F_{N} (M_{N} \leqslant  m) \\
&= F^{N} (m)
\label{eq:cdf}
\end{align}
By differentiating \Eq{cdf} we find the probability density function (PDF) of the distribution,
\begin{align}
\Phi (M_{\mathrm{max}} = m;\, N) &= N F'(m) [ F(m) ]^{N-1} \\
&= N f(m) [ F(m) ]^{N-1} \;,
\label{eq:pdf}
\end{align}
where $f(m)$ is the PDF of the original distribution ($f(m) = dF(m)/dm$), and
$\Phi (M_{\mathrm{max}} = m;\, N)$ is the \textit{exact} extreme value PDF for $N$ observations drawn from the known distribution $F(m)$.
We apply this general result to the case of the halo mass function, where $n(M)$ is the number density of haloes of mass $M$, and derive $f(M)$ and $F(M)$, for haloes in a fixed fraction of the sky, $f_{\mathrm{sky}}$, between redshifts $z_{\mathrm{min}}$ and $z_{\mathrm{max}}$,
\begin{align}
f(m) &= \frac{f_{\mathrm{sky}}}{n_{\mathrm{tot}}} \left [ \int_{z_{\mathrm{min}}}^{z_{\mathrm{max}}} dz \frac{dV}{dz} \frac{dn(m,z)}{dm} \right ] \\
F(m) &= \frac{f_{\mathrm{sky}}}{n_{\mathrm{tot}}} \left [ \int_{z_{\mathrm{min}}}^{z_{\mathrm{max}}} \int^{m}_{- \infty}  dz \, dM \frac{dV}{dz} \frac{dn(M,z)}{dM} \right ] \;,
\end{align}
where $n_{\mathrm{tot}}$ is a normalisation factor giving the total (co-moving) number of haloes,
\begin{equation}
n_{\mathrm{tot}} = f_{\mathrm{sky}} \left [ \int_{z_{\mathrm{min}}}^{z_{\mathrm{max}}} \int^{\infty}_{- \infty}  dz \, dM \frac{dV}{dz} \frac{dn(M,z)}{dM} \right ] \;.
\end{equation}

The halo mass function is the only ingredient of \Eq{pdf}; we adopt the \cite{behrooziAverageStarFormation2013} fit as our fiducial mass function, as used in \cite{lovellExtremeValueStatistics2023}, since it is defined across the full redshift range considered here.
Mass functions are computed with the \textsc{hmf} package \citep{murrayHMFcalcOnlineTool2013}, using a matter power spectrum derived from a \textsc{camb} transfer function \citep{lewisEfficientComputationCosmic2000a}.
The extrapolation of that transfer function to scales beyond those tabulated by \textsc{camb} is a significant source of uncertainty at the highest redshifts, where the most massive halo in a small field is itself of relatively low mass; we quantify this in \app{hmf_transfer}.
In \app{hmf_comparison} we compare against the \cite{yungAreUltrahighredshiftGalaxies2024} calibrated fit to the GUREFT $N$-body simulations over $z = 6$--$19$, and show that the two agree to within $\sim 10\%$ over the range where both are valid.

The prediction is therefore defined for a given volume, set by the survey area through $f_{\mathrm{sky}}$ and by the redshift interval $[z_{\mathrm{min}}, z_{\mathrm{max}}]$.
We evaluate \Eq{pdf} in contiguous redshift shells of width $\Delta z = 0.2$ across $0 < z < 28$, integrating the mass function on a sub-grid of $\delta z = 0.02$ within each shell; all quantities plotted against redshift in this paper are therefore the extreme value distribution of the corresponding shell, and not a cumulative prediction over all redshifts above $z$.
It is impractical and unnecessary to integrate between infinite endpoints in mass; we tabulate the mass function between $10^{3.5}$ and $10^{17} \, \mathrm{M_{\odot}}$ with a resolution of $0.01$ dex, and evaluate the extreme value distribution over the range $10^{4.5}$--$10^{17} \, \mathrm{M_{\odot}}$.
\cite{lovellExtremeValueStatistics2023} confirm that these numerical choices do not affect the results.
We also ignore clustering, which introduces cosmic variance in the number of haloes within a survey and lowers the predicted maximum mass \citep{jespersenSignificanceRareObjects2025}, so ignoring it is conservative.
Cosmic variance is largest for a single contiguous field and is reduced when the same area is split across separate fields; assuming a contiguous geometry would be a liberal assumption.

\subsection{From halo mass to stellar mass}
\label{sec:stellar_content}

Given the above formalism we can predict the full PDF of the most massive halo in a given survey.
To convert this to a PDF for the most massive galaxy (in terms of stellar mass) we must account for both the baryon fraction, $f_{\mathrm{b}}$, and the stellar fraction, $f_{\star}$, defined here as the fraction of the halo baryons that have been converted into stars,
\begin{equation}
M_{\star} = f_{\mathrm{b}} \, f_{\star} \, M_{\mathrm{DM}} \;.
\label{eq:mstar}
\end{equation}
The baryon fraction is set by our assumed cosmology \citep[$f_{\mathrm{b}} = 0.16$;][]{planckcollaborationPlanck2015Results2016}.

The stellar fraction is dependent on the astrophysics that converts cold gas into stars.
A conservative upper limit is to assume \textit{all} baryons are converted into stars, $f_{\star} = 1$, and simply multiply the halo PDF by the product of the baryon and stellar fractions,
\begin{equation}
\Phi(M_{\mathrm{*}}) = \Phi(M_{\mathrm{DM}}) \, f_{\mathrm{b}} \, f_{\star} \;.
\end{equation}
In reality, measurements of the stellar fraction suggest much lower values, particularly in the most massive haloes \citep[\textit{e.g.}][]{giodiniStellarTotalBaryon2009,behrooziComprehensiveAnalysisUncertainties2010}.
To account for this, we assume a lognormal distribution of $f_{\star}$, truncated to the physical range $0 \leqslant f_{\star} \leqslant 1$,
\begin{align}
    \mathrm{ln} \, f_{\star} \sim N(\mu, \sigma^2) \;,
\end{align}
with $\mu = -2$ and $\sigma = 1$, giving a median stellar fraction of $f_{\star} = e^{-2} \simeq 0.14$, and a median stellar-to-halo mass ratio $f_{\mathrm{b}} f_{\star} \simeq 0.02$.
This simple model ignores the dependence of the stellar fraction on redshift and halo mass, but incorporates the range of values inferred from simple halo models \citep{tacchellaPhysicalModel02013,tacchellaRedshiftindependentEfficiencyModel2018} and observations \citep[\textit{e.g.}][]{harikaneEvolutionStellartoHaloMass2016,harikaneGOLDRUSHIIClustering2018,stefanonGalaxyStellarMass2021}.

Since $f_{\star}$ is a distribution rather than a fixed value, the stellar mass PDF is obtained by convolving the halo extreme value PDF with the distribution of $f_{\star}$, scaled by the baryon fraction.\footnote{\cite{lovellExtremeValueStatistics2023} evaluated the same product by Monte Carlo sampling; the convolution gives the same result without the sampling noise.}
In doing so we assume that $f_{\star}$ is independent of halo mass.
This is a reasonable approximation at high redshift \citep{somervilleDensityModulatedStar2025}, but breaks down in the cluster regime at low redshift where $f_{\star}$ falls systematically with halo mass; we therefore restrict our predictions to $z > 2$.
Additionally, we treat the distribution of $f_{\star}$ as uncertainty in the stellar fraction common to all haloes in the volume, rather than as a scatter between haloes; the most massive halo therefore also hosts the most massive galaxy.
We examine the impact of galaxy-to-galaxy variability in \sec{mass_to_flux} and \app{burstiness}.
We also present upper $3\sigma$ limits based on assuming $f_{\star} = 1$ and applying directly to the halo EVS PDF, which can be interpreted as conservative upper limits.
For a given emission model the flux of a galaxy is proportional to its stellar mass (\sec{mass_to_flux}), so at each redshift the flux distribution is the stellar mass distribution shifted by a constant in log space.
The halo mass function and the distribution of $f_{\star}$ therefore set the shape and width of the prediction, and the emission model sets only its position.

In \app{stellar_mass} we re-evaluate the published stellar mass estimates from early JWST analyses, as well as new photometric candidates at extreme redshifts, in stellar mass space, updating the analysis presented in \cite{lovellExtremeValueStatistics2023}.

\subsection{Eddington bias correction}
\label{sec:eddington}

Comparisons made in terms of the halo or stellar mass must account for Eddington bias \citep{eddingtonFormulaCorrectingStatistics1913}: since the mass function falls steeply, symmetric measurement uncertainties scatter more objects up in mass than down, and the most extreme object is on average less massive than its measured value implies \citep{lovellExtremeValueStatistics2023}.
The correction scales as the square of the uncertainty, and is significant for SED-derived stellar masses; for the fluxes that form the bulk of this work, where the measurement uncertainties are far smaller, it is negligible.
A more relevant uncertainty is the photometric redshift, which is large and frequently asymmetric for many high-$z$ candidates.
Because the predicted flux declines steeply with redshift, this can bias the comparison in a similar way to the Eddington bias; we do not attempt to correct for it here, so the offsets quoted in \sec{comparison} may overstate the tension.

\section{Modelling observables}
\label{sec:obs}

The observed multiwavelength emission from a galaxy is dependent on a number of different physical components and processes acting within and outside the galaxy \citep{conroyModelingPanchromaticSpectral2013,iyerSpectralEnergyDistributions2026}.
This includes the source stars, the nebular gas in young star forming regions, the 3D distribution of dust in the interstellar medium, and the intervening neutral gas between the galaxy and the observer in the intergalactic medium.

However, the main drivers of galaxy luminosities and colours are the source stars themselves; the emission is closely linked to the age and metallicity distribution, known as the star formation and metal enrichment history (SFZH).
A number of parametric forms have been proposed to describe this distribution, usually in terms of the ages at fixed metallicity.
This 1D marginal distribution of stellar ages of all stars ever born, not just the surviving mass, is known as the star formation history (SFH).

\subsection{Star formation history modelling}

We assume a number of SFH forms in this analysis that have been proposed in the literature for modelling both nearby and high-$z$ galaxies \citep{salmonRelationStarFormation2015,cieslaSFRMMainSequence2017,whitlerStarFormationHistories2023,tacchellaStellarPopulationsGalaxies2022,dresslerEarlyResultsGLASSJWST2023,dresslerBuildingFirstGalaxies2024}.
These include a lognormal, double power law, and exponentially rising form; \cite{carnallHowMeasureGalaxy2019} compare all three, along with the exponentially declining and delayed forms more commonly applied at low redshift, and quantify the priors each imposes.
In each case we parametrise the star formation rate as a function of cosmic time, $t$, with $t = 0$ corresponding to the Big Bang, and fix the overall normalisation by the total stellar mass formed by the time of observation.

The lognormal SFH \citep{gladdersIMACSClusterBuilding2013,diemerLognormalStarFormation2017} is given by
\begin{equation}
    \psi(t) \propto \frac{1}{t} \, \mathrm{exp} \left[ - \frac{(\ln t - T_{0})^{2}}{2 \tau^{2}} \right] \,,
    \label{eq:sfh_lognormal}
\end{equation}
where $T_{0}$ sets the logarithmic peak time and $\tau$ controls the width of the distribution.
The double power law form \citep{behrooziAverageStarFormation2013,carnallInferringStarFormation2018} is
\begin{equation}
    \psi(t) \propto \left[ \left( \frac{t}{t_{\mathrm{pk}}} \right)^{\alpha} + \left( \frac{t}{t_{\mathrm{pk}}} \right)^{-\beta} \right]^{-1} \,,
    \label{eq:sfh_dpl}
\end{equation}
where $t_{\mathrm{pk}}$ sets the peak time, and $\alpha$ and $\beta$ are the falling and rising power-law indices at late and early times respectively.
The exponentially rising form, sometimes termed the inverted-$\tau$ model, is
\begin{equation}
    \psi(t) \propto \mathrm{exp}(t / \tau) \,,
    \label{eq:sfh_exp}
\end{equation}
where $\tau$ is the e-folding time.
Rising histories of this kind were introduced precisely because the declining $\tau$-models developed for local galaxies return implausibly young ages and low masses when applied at high redshift, and are supported by the observed growth of both the star formation rate and stellar mass of galaxies from $z \sim 8$ to $z \sim 3$ \citep{papovichRisingStarFormation2011,reddyCharacteristicStarFormation2012}.
At a given redshift we truncate the star formation history so that the maximum stellar age is no greater than the age of the Universe at that redshift.

\begin{figure*}
	\includegraphics[width=\textwidth]{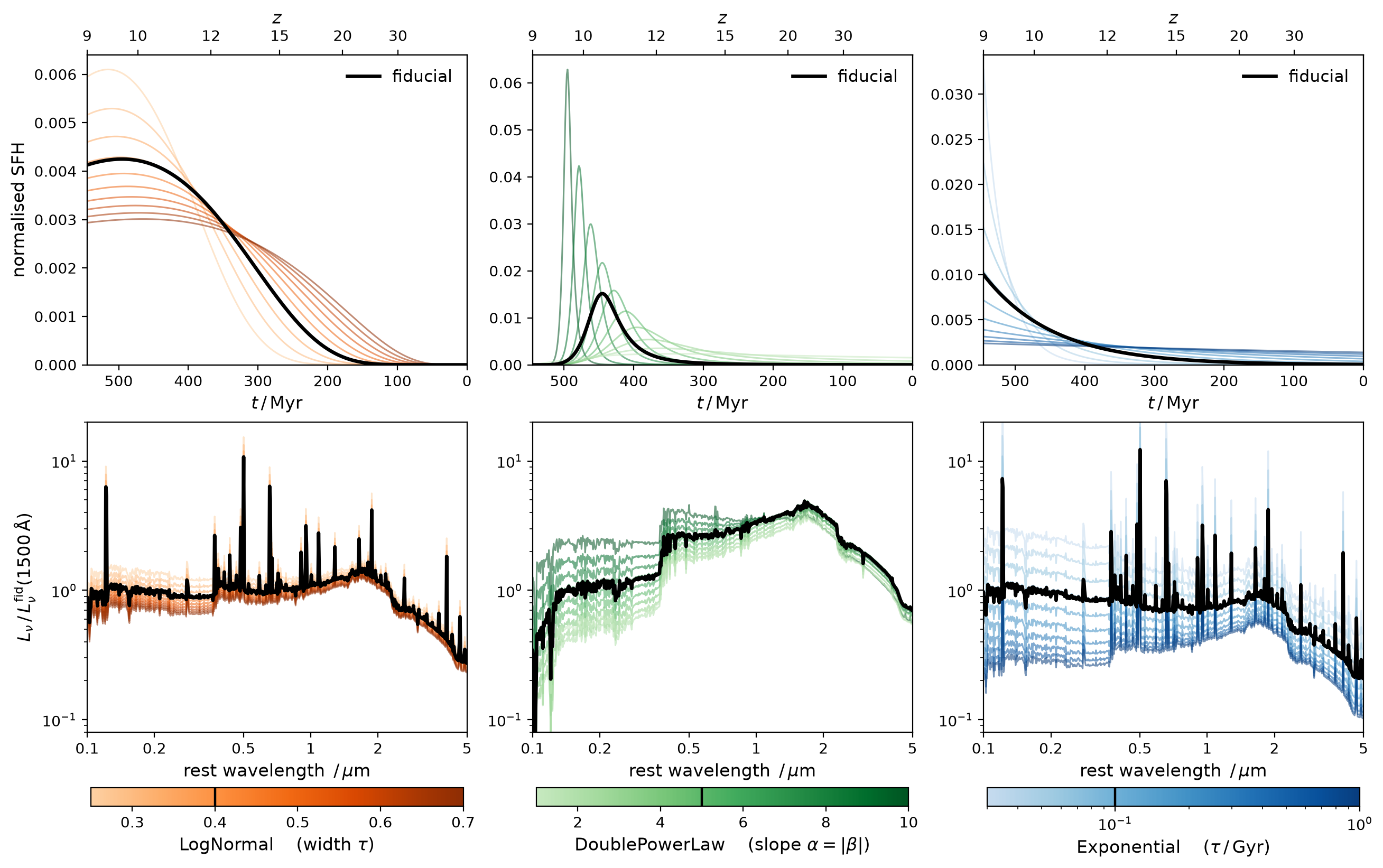}
    \caption{The lognormal (\textit{left}), double power law (\textit{centre}) and exponentially rising (\textit{right}) parametric star formation history forms adopted in this work, evaluated at $z = 9$. \textit{Top row:} the star formation histories themselves, shown as normalised mass-fraction densities against cosmic time $t$, with the corresponding redshift on the upper axis. Time runs from the epoch of observation on the left back to the Big Bang on the right, and each history is truncated at the age of the Universe at $z = 9$. \textit{Bottom row:} the corresponding rest-frame spectra, computed with the fiducial forward modelling choices. Each panel is normalised to the rest-UV (1500\,\AA) luminosity of its own fiducial model. In both rows the bold black curve is our fiducial model, while the coloured curves sweep the comparison range; the swept parameter is indicated by the colourbar, with the fiducial value marked.}
    \label{fig:sfh_models}
\end{figure*}

We choose a \textit{fiducial} parameter set for each form together with a comparison set bracketing sharply-peaked to broadly-extended histories.
Since most studies employ non-parametric SFHs, we match their qualitative properties rather than adopting their best-fit parameters directly.
For the exponential we take a fiducial e-folding time $\tau = 0.1\,\mathrm{Gyr}$ (implying $\mathrm{sSFR} \approx 10\,\mathrm{Gyr^{-1}}$ and a mass-doubling time of $\sim 70\,\mathrm{Myr}$ at $z \sim 7$), and vary $\tau$ from $0.03$ to $1\,\mathrm{Gyr}$ (rapidly rising to near-constant).
For the lognormal we adopt a fiducial width $\tau = 0.4$ with the peak set $\sim 50\,\mathrm{Myr}$ before observation (a rise time of $\sim 190\,\mathrm{Myr}$), spanning $\tau = 0.25$ (narrow, bursty) to $0.7$ (broad, extended).
For the double power law we adopt a symmetric fiducial $\alpha = \beta = 5$ peaking $\sim 100\,\mathrm{Myr}$ before observation (a $\sim 150\,\mathrm{Myr}$ rise-and-fall), and vary the slopes from $\alpha = \beta = 10$ (a sharp $\sim 50\,\mathrm{Myr}$ episode) to $\alpha = \beta = 1$ (an extended plateau); this form additionally captures the rapid post-peak quenching inferred for the first spectroscopically-confirmed quiescent galaxies at $z > 4$ \citep{carnallMassiveQuiescentGalaxy2023a,weibelRUBIESRevealsMassive2025}.
These fiducial models and their comparison ranges are shown in \fig{sfh_models}.

\subsection{Forward modelling with Synthesizer}

We use \synthesizer{}\footnote{\href{https://synthesizer-project.github.io/synthesizer/}{https://synthesizer-project.github.io/synthesizer/}} \citep{lovellSynthesizerSoftwarePackage2025,roperSynthesizerSyntheticObservables2026}, a modular Python framework for the forward modelling of galaxy spectral energy distributions (SEDs), photometric fluxes, emission lines and resolved images.
Galaxies in \synthesizer{} may be represented either as discrete distributions of star particles, drawn from cosmological simulations, or via parametric descriptions of the star formation and metal enrichment history (SFZH); we adopt the latter representation throughout this work.
At the core of \synthesizer{} is the \grid{} object, which encapsulates a pre-computed table of simple stellar population (SSP) spectra on a regular grid of stellar age and metallicity, drawn from a chosen SPS model.
\grids{} may optionally include a self-consistent reprocessing of the ionising continuum, and the resulting nebular line and continuum emission, computed with a photoionisation code \citep{vijayanStellarPhotoionisationModelling2026a}.
Emission components, including the stellar continuum, nebular emission, dust attenuation and re-emission, and any AGN contribution, are composed through the \emodel{} framework, which represents each component as a node in a flexible directed graph of transformations and combinations.
Units are propagated consistently throughout the calculation via the \unyt{} library \citep{goldbaumUnytHandleManipulate2018}.

For each parametric SFZH considered in this work we compute the rest-frame stellar SED by convolving the SFZH with the SSP spectra on the chosen \grid{}, redshift the resulting spectrum to the target redshift, and integrate against the relevant JWST/NIRCam transmission curve obtained from SVO \citep{rodrigoSVOFilterProfile2020,rodrigoSVOFilterProfile2012} to obtain the observed-frame photometric flux in each band.

We use BPASS version 2.2.1 as our fiducial SPS model \citep{eldridgeBinaryPopulationSpectral2017,stanwayReevaluatingOldStellar2018,byrneDependenceTheoreticalSynthetic2022}, which incorporates the effect of binary stellar evolution on the integrated emission from a simple stellar population (SSP).
We assume the inclusion of binaries, and a \cite{chabrierGalacticStellarSubstellar2003} IMF between $0.1$ and $300 \, \mathrm{M_{\odot}}$.
The grid is tabulated on $51$ ages, logarithmically spaced in $0.1$ dex intervals over $10^{6} < \mathrm{age/yr} < 10^{11}$, and $13$ metallicities spanning $10^{-5} < Z < 0.04$; we take $Z = 0.01$ as our fiducial value, and explore the dependence on this choice in \sec{systematics}.
The ionising continuum of each SSP is reprocessed with \textsc{cloudy} c23.01 \citep{gunasekera2025ReleaseCloudy2025a}, assuming a spherical, constant-density nebula with hydrogen density $n_{\mathrm{H}} = 10^{3} \, \mathrm{cm^{-3}}$ and an ionisation parameter of $U = 0.01$, referenced to a $1 \, \mathrm{Myr}$, $Z = 0.01$ SSP and scaled with the ionising output elsewhere on the grid.
We assume no escape of ionising photons ($f_{\mathrm{esc}} = 0$), so that the nebular contribution is maximal; the comparison grids used in \sec{systematics} are processed identically.

We ignore the impact of dust in this analysis.
Dust acts to attenuate the UV-optical emission, re-emitting the energy in the infrared and sub-mm.
Since we concentrate here on predictions for JWST, which probes the rest-frame UV-optical at high redshift, the effect of dust will be to reduce the flux.
By ignoring dust we are therefore making a \textit{conservative upper limit} on the emission, which is appropriate for this analysis studying the brightest possible objects in the Universe. 
Dust-free, or strongly dust-depleted, galaxies have also been proposed at high redshift, both on theoretical grounds invoking radiation-driven outflows that evacuate dust from young, compact starbursts \citep{ferraraSuperearlyJWSTGalaxies2024,nathDustfreeStarburstGalaxies2023}, and observationally, from UV continuum slopes that become bluer towards high redshift and at $z > 10$ approach the intrinsic limit of dust-free stellar populations \citep{cullenUltravioletContinuumSlopes2023,toppingUVContinuumSlopes2024a,cullenUltravioletContinuumSlopes2024}, so this is also not an unreasonable assumption in and of itself.

\begin{table}
	\centering
	\caption{Fiducial modeling choices, and the description of any variations in the text.}
	\label{tab:fiducial}
	\setlength{\tabcolsep}{4pt}
	\begin{tabular}{l>{\raggedright\arraybackslash}p{0.47\columnwidth}l}
		\hline
		Ingredient & Fiducial choice & Varied in \\
		\hline
		Cosmology & \citealt{planckcollaborationPlanck2015Results2016} & -- \\
		Halo mass function & \citealt{behrooziAverageStarFormation2013}; \textsc{camb} transfer function with \citealt{eisensteinBaryonicFeaturesMatter1998} extrapolation & App.~\ref{sec:hmf_transfer}, \ref{sec:hmf_comparison} \\
		Stellar fraction & $\ln f_{\star} \sim \mathcal{N}(-2, 1)$, $f_{\star} \leqslant 1$ & $f_{\star} = 1$ limit \\
		SFH & Exponentially rising, $\tau = 0.1$ Gyr & Sec.~\ref{sec:systematics} \\
		SPS model & \textsc{bpass} v2.2.1, with binaries & Sec.~\ref{sec:systematics} \\
		IMF & \citealt{chabrierGalacticStellarSubstellar2003}, $0.1$--$300\,\mathrm{M_{\odot}}$ & Sec.~\ref{sec:systematics} \\
		Metallicity & $Z = 0.01$ & Sec.~\ref{sec:systematics} \\
		Nebular emission & \textsc{cloudy} c23.01, $n_{\mathrm{H}} = 10^{3}\,\mathrm{cm^{-3}}$, $U = 0.01$, $f_{\mathrm{esc}} = 0$ & Sec.~\ref{sec:systematics} \\
		Burstiness & None & App.~\ref{sec:burstiness} \\
		\hline
	\end{tabular}
\end{table}

\subsection{From stellar mass to flux}
\label{sec:mass_to_flux}

Each SSP spectrum on the \grid{} is tabulated per unit initial stellar mass, and the convolution with the SFZH described above is linear in that mass.
The observed flux density of a galaxy of stellar mass $M_{\star}$ at redshift $z$, in band $X$, is therefore simply proportional to its stellar mass,
\begin{equation}
    F^{X}_{\nu}(M_{\star}, z) = M_{\star} \, \mathcal{F}^{X}_{\nu}(z; \theta) \,,
    \label{eq:flux_linear}
\end{equation}
where $\mathcal{F}^{X}_{\nu}$ is the flux per unit stellar mass and $\theta$ collects the emission model choices, such as the assumed SFH, the metallicity, the SPS model, the IMF and the treatment of nebular emission.
Writing the rest-frame luminosity density per unit stellar mass formed as
\begin{equation}
    l_{\nu}(\lambda; z, \theta) = \int_{0}^{t(z)} \mathrm{d}t \; \psi(t) \, l^{\mathrm{SSP}}_{\nu} \left[ \lambda, \, t(z) - t, \, Z \right] \,,
    \label{eq:lnu}
\end{equation}
where $\lambda$ is the rest-frame wavelength, $t(z)$ the age of the Universe at redshift $z$, and $\psi(t)$ the assumed star formation history, normalised to unit mass formed by the time of observation.
$l^{\mathrm{SSP}}_{\nu}$ is the tabulated \grid{} of SSP luminosity densities per unit initial stellar mass, as a function of wavelength, stellar age and metallicity $Z$.
We assume a single metallicity for the whole population, so that the SFZH separates into the product of $\psi(t)$ and a delta function in $Z$, and \Eq{lnu} is an integral over age alone.
The band flux is then the integral over the filter transmission $T_{X}$,
\begin{equation}
    \mathcal{F}^{X}_{\nu}(z; \theta) = \frac{(1+z)}{4 \pi d^{2}_{\mathrm{L}}(z)} \,
    \frac{\int \mathrm{d}\lambda \; \lambda^{-1} \, T_{X}(\lambda) \, e^{-\tau_{\mathrm{IGM}}(\lambda, z)} \, l_{\nu} \! \left[ \lambda / (1+z) \right]}
         {\int \mathrm{d}\lambda \; \lambda^{-1} \, T_{X}(\lambda)} \,,
    \label{eq:fnu_band}
\end{equation}
where the integrals now run over observed-frame wavelength, $d_{\mathrm{L}}$ is the luminosity distance and $\tau_{\mathrm{IGM}}$ the \cite{madauHighredshiftGalaxiesHubble1996} attenuation due to the intergalactic medium (IGM).
The redshift dependence of $\mathcal{F}^{X}_{\nu}$ is twofold: dimming and shifting of the SED through $d_{\mathrm{L}}$ and $(1+z)$, and the truncation of the SFH at the age of the Universe $t(z)$, which makes the stellar population intrinsically younger and bluer towards high redshift.
$M_{\star}$ is defined as the total mass of stars \textit{formed}, neglecting mass return to the ISM by evolved stars; for the young populations relevant here at high redshift this is a small effect, and one that acts in the same conservative direction as our neglect of dust.

\Eq{flux_linear} is a monotonic map at fixed $z$ and $\theta$, so the extreme value distribution of the flux follows from that of the stellar mass; its percentiles map one-to-one onto those of the stellar mass distribution.
We therefore evaluate all confidence intervals directly in flux space.
Combining with the stellar mass distribution of \sec{stellar_content} gives
\begin{equation}
    \begin{split}
    \Phi \left( \mathrm{log_{10}} F \right) & = \int \mathrm{d} \, \mathrm{log_{10}} f_{\star} \; p(\mathrm{log_{10}} f_{\star}) \\
    & \quad \times \; \Phi_{\mathrm{h}} \! \left( \mathrm{log_{10}} F - \mathrm{log_{10}} \mathcal{F}^{X}_{\nu} - \mathrm{log_{10}} f_{\mathrm{b}} f_{\star} \right) \,,
    \end{split}
    \label{eq:flux_full}
\end{equation}
where $\Phi_{\mathrm{h}}$ is the extreme value PDF of the halo mass distribution, \Eq{pdf}, and $p(\mathrm{log_{10}} f_{\star})$ the stellar fraction distribution.\footnote{The dependence of every term on redshift has been suppressed for clarity.}
The conservative $f_{\star} = 1$ limits shown throughout simply replace $p(\mathrm{log_{10}} f_{\star})$ with a delta function at $f_{\star} = 1$.

\Eq{flux_full} describes the galaxy hosted by the most massive halo.
Since galaxies in haloes of similar mass differ in their recent star formation, a less massive halo caught in a burst may instead host the brightest galaxy.
In \app{burstiness} we quantify this by superimposing stochastic fluctuations on each SFH, with an amplitude ($0.24$ dex) and correlation time ($50$ Myr) measured from high-redshift simulations \citep{pallottiniStochasticStarFormation2023}, and apply the resulting galaxy-to-galaxy scatter in flux to the halo population before computing the extreme value distribution.
This raises the predicted flux of the brightest galaxy by $\lesssim 0.1$ dex for each SFH form, at both the median and the $+3\sigma$ contour, small compared with the systematics quantified in \sec{systematics}, and we neglect it in the main analysis.
Substantially burstier histories raise the prediction by $\sim 0.2$--$0.45$ dex, but imply a scatter in stellar mass at fixed halo mass of $\sim 0.3$ dex, above that typically inferred \citep{behrooziComprehensiveAnalysisUncertainties2010,enriquez-vargasEmpiricalEstimatesHow2026}.

Our fiducial modelling choices are summarised in \Tab{fiducial}.

\begin{figure*}
	\includegraphics[width=\textwidth]{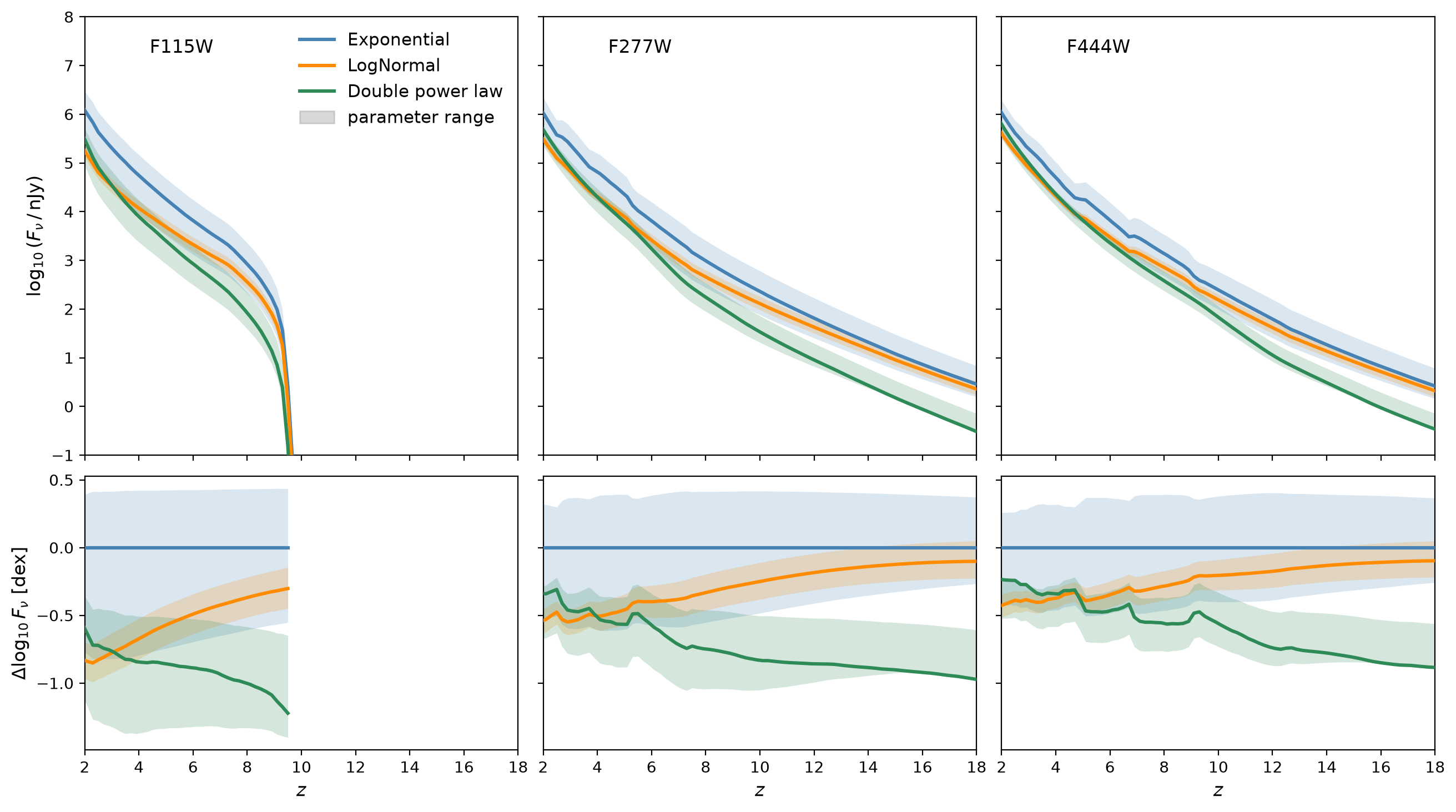}
    \caption{Sensitivity of the EVS flux prediction to the assumed parametric star formation history, in three JWST/NIRCam bands (columns), for the $0.28\,\mathrm{deg^{2}}$ COSMOS-Web area. Curves show the median EVS flux of the brightest galaxy for each of the three SFH forms of \Fig{sfh_models}: exponentially rising (blue), lognormal (orange) and double power law (green).
    Solid lines are the fiducial parameters of each form, and the shaded bands span the parameter ranges swept in \Fig{sfh_models}. All other ingredients are held at their fiducial values.
    The top row shows the median flux, and the bottom row its difference (in dex) from the fiducial exponentially rising SFH, for direct comparison with \Fig{sps_imf}.
    The F115W curves drop out at $z \simeq 9.6$ as the Lyman break passes out of the filter.
    The small bumps in the medians are due to emission lines passing in and out of each filter with redshift.
    }
    \label{fig:evs_tau}
\end{figure*}

\section{Results}
\label{sec:results}

\subsection{Impact of forward modelling assumptions}
\label{sec:systematics}

The prediction of \sec{mass_to_flux} depends on the emission model only through the flux per unit stellar mass, $\mathcal{F}^{X}_{\nu}$. 
When imposed uniformly across the stellar population, each emission-model choice acts as a redshift-dependent shift in $\mathcal{F}^{X}_{\nu}$.
We quantify each in turn, varying one ingredient at a time about the fiducial model while holding the others fixed, and comparing the median EVS flux for the $0.28\,\mathrm{deg^{2}}$ COSMOS-Web area in the F115W, F277W and F444W NIRCam bands.

We begin with the star formation history, sweeping each of the three parametric forms of \Fig{sfh_models} across the parameter ranges given in \sec{obs} (\Fig{evs_tau}).
Varying the parameters within a single form produces a spread in the predicted flux at fixed redshift, but the systematic differences between the forms are larger still.
The double power law with a late peak age and shallow $\alpha$ and $\beta$ slopes gives the faintest prediction, lying $\sim 0.9\,\mathrm{dex}$ below the exponentially rising form by $z \sim 18$.
This double power law form therefore provides the most liberal theoretical limit, and is the choice against which an observed candidate is most likely to appear in tension.

\begin{figure*}
	\includegraphics[width=\textwidth]{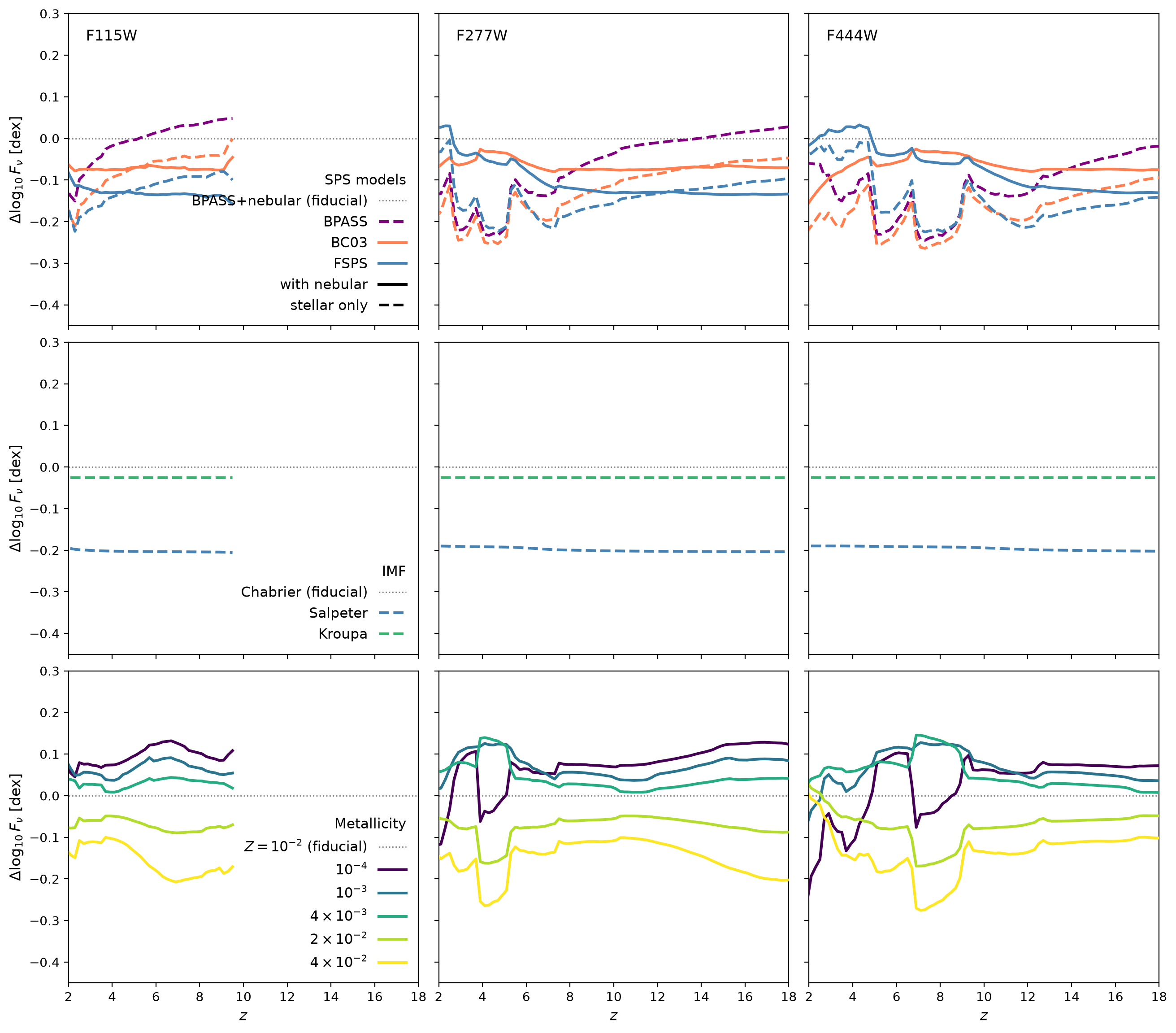}
    \caption{Sensitivity of the EVS flux prediction to the assumed SPS model (\textit{top row}), IMF (\textit{middle row}) and stellar metallicity (\textit{bottom row}), for the fiducial exponentially rising SFH, in three JWST/NIRCam bands (columns).
    Each panel shows the median EVS flux of the brightest galaxy as a difference (in dex) relative to a fiducial model (grey dotted line): \textsc{bpass} including nebular emission for the SPS row, a Chabrier IMF for the IMF row, and $Z = 10^{-2}$ for the metallicity row.
    Solid curves include nebular emission, dashed curves are stellar emission only. \textit{Top:} \textsc{bc03} and \textsc{fsps}, each shown with and without nebular emission, alongside the stellar-only \textsc{bpass} model. \textit{Middle:} the Salpeter and Kroupa IMFs relative to Chabrier, all within the \textsc{bc03} intrinsic model. \textit{Bottom:} metallicities spanning $10^{-4} \leq Z \leq 4 \times 10^{-2}$, coloured by $Z$.}
    \label{fig:sps_imf}
\end{figure*}

We next vary the SPS model, holding the fiducial exponentially rising SFH fixed (\Fig{sps_imf}, top row).
Alongside the fiducial \textsc{bpass} grid we use \textsc{bc03} \citep[the 2016 revision, using the MILES spectral library;][]{bruzualStellarPopulationSynthesis2003} and \textsc{fsps} v3.2 \citep[MIST isochrones and MILES;][]{conroyPropagationUncertaintiesStellar2009}, the latter with a \cite{chabrierGalacticStellarSubstellar2003} IMF between $0.08$ and $100$ $\mathrm{M_{\odot}}$, close to the mass range of the \textsc{bc03} grid, each in both a stellar-only and a \textsc{cloudy}-processed version so that the SPS and nebular contributions can be separated.
The sensitivity to the SPS model is modest, the predictions lying within $\lesssim 0.15\,\mathrm{dex}$ of the fiducial across the full redshift range, and at $z \gtrsim 5$ the fiducial \textsc{bpass} model gives the brightest prediction of the three.
The treatment of nebular emission matters considerably more: the fiducial SFH is young and rising, so the ionising photon budget is large, and including photoionisation raises the predicted flux by up to $\sim 0.26\,\mathrm{dex}$ relative to the stellar continuum alone.
The structure in the flux curves traces strong rest-frame optical lines moving in and out of each filter.

Varying the IMF between the \citet{chabrierGalacticStellarSubstellar2003}, \citet{salpeterLuminosityFunctionStellar1955} and \citet{kroupaVariationInitialMass2001} forms, within the \textsc{bc03} model and over a common $0.1$--$100 \, \mathrm{M_{\odot}}$ mass range so that only the slope varies, gives shifts slightly larger than those introduced by the SPS model (\Fig{sps_imf}, middle row).
The Salpeter IMF, with its greater weight at low stellar masses, gives the largest offset, lying $\sim 0.20\,\mathrm{dex}$ below the Chabrier and Kroupa predictions.

Finally we vary the stellar metallicity between $Z = 10^{-4}$ and $Z = 4 \times 10^{-2}$, spanning $\sim 0.007$--$3 \, Z_{\odot}$ and reaching the upper limit of the \textsc{bpass} grid, holding the SFH and SPS model fixed at their fiducial values (\Fig{sps_imf}, bottom row).
The predicted flux decreases monotonically with increasing metallicity over most of the redshift range, as expected from the reddening of the stellar continuum and the reduced ionising photon output of metal-rich populations; relative to the fiducial $Z = 10^{-2}$ the extremes of the range lie within $+0.13$ and $-0.21\,\mathrm{dex}$ at $z \sim 10$--$18$, a total spread of $\sim 0.3\,\mathrm{dex}$.
The exception is the most metal-poor model at $z \lesssim 9$, which falls \textit{below} the fiducial in the redder bands: the rest-frame optical lines that dominate the F277W and F444W fluxes at these redshifts are strongly suppressed at $Z = 10^{-4}$, and the structure in all of the curves tracks these lines moving through the filters.

In summary, the choice of SFH form dominates, at $\sim 0.9\,\mathrm{dex}$ by $z \sim 18$; the metallicity ($\sim 0.3\,\mathrm{dex}$), the nebular treatment ($\sim 0.26\,\mathrm{dex}$), the IMF ($\sim 0.20\,\mathrm{dex}$) and the SPS model ($\lesssim 0.15\,\mathrm{dex}$) each have a smaller effect, and act in both directions rather than coherently; galaxy-to-galaxy burstiness raises the prediction by a further $\lesssim 0.1\,\mathrm{dex}$ (\app{burstiness}).

\subsection{Comparison with observed candidates}
\label{sec:comparison}

We now apply the fiducial model to four published samples of luminous high-redshift candidates, spanning two orders of magnitude in survey area.
The first sample, the $z \gtrsim 9$ sources identified in the $0.28 \, \mathrm{deg^{2}}$ COSMOS-Web survey \citep{caseyCOSMOSWebIntrinsicallyLuminous2024} extends to $z \simeq 14$.
Three recent searches push further: the $17.6 \, \mathrm{arcmin^{2}}$ MIDIS+NGDEEP overlap \citep{perez-gonzalezRiseGalacticEmpire2025a}, the $615 \, \mathrm{arcmin^{2}}$ ASTRODEEP-JWST fields \citep{castellanoPushingJWSTExtremes2025a}, and a $> 0.6 \, \mathrm{deg^{2}}$ search across more than 150 sight-lines \citep{mcleodSearchFirstGalaxies2026a}.
Because the extreme value distribution depends on the volume surveyed, each sample is compared against the prediction for its own area, and the contours are therefore not interchangeable between samples.
We restrict the comparison to $z > 2$, since at lower redshift the most massive haloes in these volumes are cluster-scale, where both the baryon fraction and the stellar fraction evolve significantly with mass and our mass-independent treatment of \sec{stellar_content} breaks down.
The advantage of proceeding this way is that the comparison is made directly in the observational plane: the quantity being compared is the measured broadband flux, rather than a stellar or halo mass inferred by inverse modelling, so the modelling uncertainty is carried entirely by the prediction, where it can be enumerated explicitly, as in \sec{systematics} above.
We quote the position of each candidate within the extreme value distribution both as a Gaussian-equivalent significance and as a probability to exceed, $\mathrm{PTE} = P(F_{\nu}^{\mathrm{max}} > F_{\nu}^{\mathrm{obs}})$, evaluated in the redshift shell containing the candidate.
Both are evaluated at the published redshift of each candidate, and are therefore conditional on it; we do not marginalise over the photometric redshift distributions (see \sec{eddington}).

\subsubsection{The COSMOS-Web sample}
\label{sec:lowz}

 \begin{figure*}
	\includegraphics[width=\textwidth]{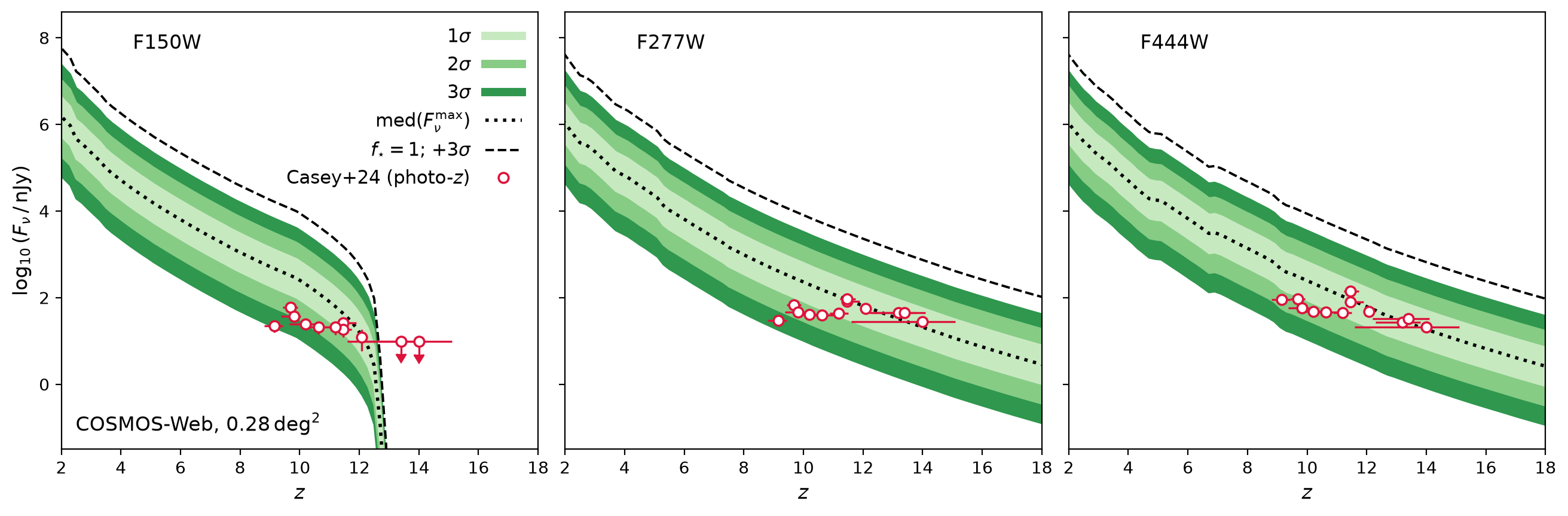}
    \caption{
        EVS flux predictions for the brightest galaxy expected in the $0.28\,\mathrm{deg^{2}}$ COSMOS-Web survey area, in three JWST/NIRCam bands (columns), compared against the luminous $z \gtrsim 9$ candidates of \citet{caseyCOSMOSWebIntrinsicallyLuminous2024} (red points) using their aperture-based photometry and \textsc{bagpipes} redshifts; downward arrows denote $1\sigma$ upper limits.
        Here and in \Fig{evs_f356w}, open and filled points denote photometric and spectroscopic redshifts, respectively.
        Shaded regions show the $1\sigma$, $2\sigma$ and $3\sigma$ confidence intervals of the EVS flux distribution (assuming the fiducial forward modelling choices); the dotted line shows the median, and the dashed line the conservative $+3\sigma$ limit obtained by assuming all halo baryons are converted into stars ($f_{\star} = 1$).
    }
    \label{fig:evs_confs}
\end{figure*}

The COSMOS-Web sample is shown in \Fig{evs_confs}.
The structure visible in the contours is a property of the filters rather than the model: the steep decline in F150W at $z \simeq 12.6$ is the Lyman break passing through the filter, and the steps in F277W and F444W are strong rest-frame optical lines entering and leaving the bands.
The candidates cluster about the median prediction for the area from which they were drawn.
The only exception is in the F150W band for \cite{caseyCOSMOSWebIntrinsicallyLuminous2024}, where there are flux upper limits beyond the Lyman-break; since we do not forward model an escape fraction of ionising photons, we do not predict any galaxies with flux beyond this limit, but given these are upper limits there is no implied tension here.
The extreme value distribution describes the brightest galaxy in each redshift shell, so only its upper contours test a candidate.
Where a shell contains several candidates, only the brightest is a draw from this distribution.
The brightest galaxy in a shell may also be absent from a sample, having been missed by the selection or assigned a different redshift.
Candidates near or below the lower contours are therefore not in tension with the model.
The extreme value distribution is broad, spanning $2.5$ dex between its $\pm 3\sigma$ contours, and every measurement remains at least $1.0$ dex below the $+3\sigma$ contour, and at least $1.4$ dex below the $f_{\star} = 1$ limit, the conservative bound obtained by converting every halo baryon into stars.
COSMOS-Web has therefore not detected an object as bright as the extreme value formalism permits.

The margin to the $3\sigma$ contour is, however, comparable to the modelling systematics quantified above.
The dominant one, the choice of SFH form, spans $0.7$--$1.2$ dex, depending on the band, at the redshifts of these candidates: replacing the fiducial exponential with the most liberal double power law of \Fig{sfh_models} (a peak age of $200\,\mathrm{Myr}$ and $\alpha = |\beta| = 1$) lowers the contours by this amount, bringing the brightest COSMOS-Web candidate onto the $+3\sigma$ contour.
The SPS model, IMF and metallicity each contribute $\lesssim 0.3$ dex and act in both directions, and are subdominant to this.
We conclude that there is no tension in our fiducial model, or across the explored parameter space, except for the choice of star formation history: under the most liberal parameterisation the brightest candidate reaches the $3\sigma$ bound.

\subsubsection{Searches beyond $z \sim 13$}
\label{sec:highz}

\begin{figure*}
	\includegraphics[width=\textwidth]{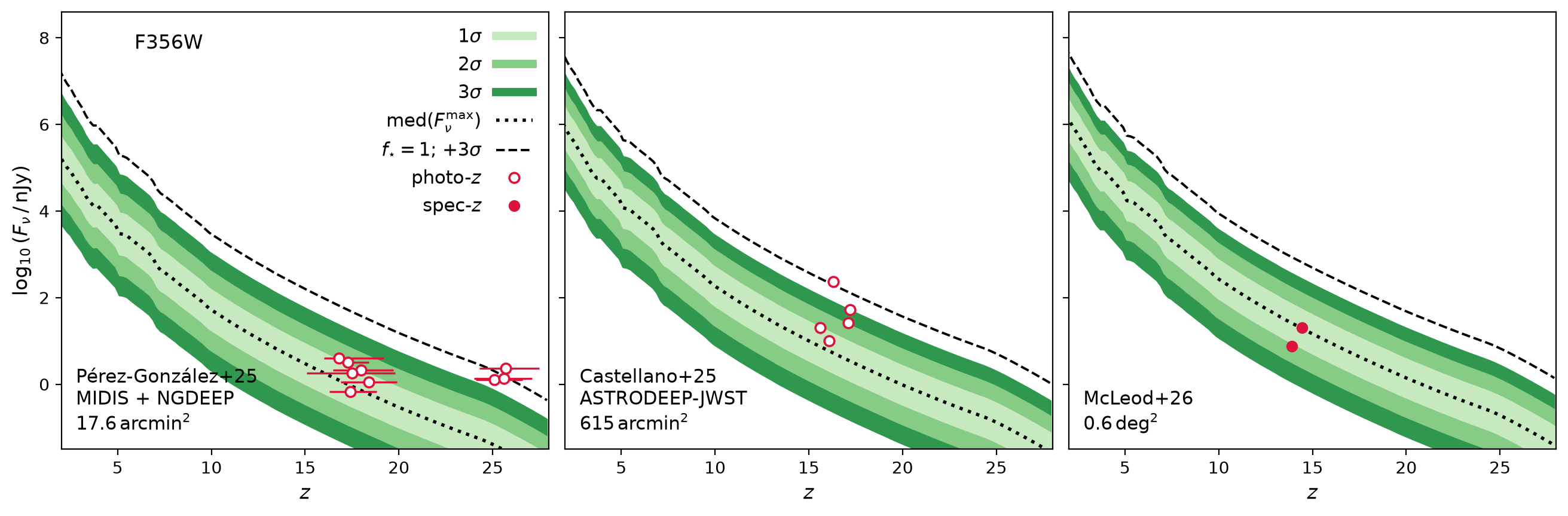}
    \caption{EVS flux predictions in F356W, compared against three searches for galaxies at $z > 13$. Contours, median and $f_{\star} = 1$ limit are defined as they are in \protect\fig{evs_confs}, and open and filled points denote photometric and spectroscopic redshifts, respectively.
    \textit{Left:} the $z \sim 17$ and $z \sim 25$ dropouts of \protect\cite{perez-gonzalezRiseGalacticEmpire2025a} over the $17.6 \, \mathrm{arcmin^{2}}$ common to MIDIS and NGDEEP, using their \textsc{eazy} redshifts and F356W magnitudes. \textit{Centre:} the five reference F200W dropouts of \protect\cite{castellanoPushingJWSTExtremes2025a} over the $615 \, \mathrm{arcmin^{2}}$ ASTRODEEP-JWST area, plotted at their best-fit $z > 10$ solution. \textit{Right:} the two objects from the $> 0.6 \, \mathrm{deg^{2}}$ search of \protect\cite{mcleodSearchFirstGalaxies2026a} with published F356W photometry, both spectroscopically confirmed: JADES-GS-z14-1 \protect\citep{carnianiSpectroscopicConfirmationTwo2024} and MoM-z14 \protect\citep{naiduCosmicMiracleRemarkably2026}.
    }
    \label{fig:evs_f356w}
\end{figure*}

The three searches reaching beyond $z \sim 13$ are compared in \Fig{evs_f356w}, in F356W: it is the one band common to all three selections, and remains above the Lyman break out to $z \simeq 28$.
The two spectroscopically confirmed objects from \citet{mcleodSearchFirstGalaxies2026a} lie at $-1.2\sigma$ and $+0.1\sigma$ of the median for their area ($\mathrm{PTE} = 0.88$ and $0.48$): the brightest galaxies known at $z \simeq 14$ are entirely typical of what the halo mass function permits, and place no strain on the formalism.
Tension appears only in the photometric samples.
The $z \sim 17$ candidates of \citet{perez-gonzalezRiseGalacticEmpire2025a} span $-0.4\sigma$ to $+1.1\sigma$ ($\mathrm{PTE} = 0.64$ to $0.13$), but their three $z \sim 25$ candidates sit at $+3.6\sigma$ to $+4.4\sigma$ ($\mathrm{PTE} = 2 \times 10^{-4}$ to $5 \times 10^{-6}$) from their median photometric redshift solution.
The five \citet{castellanoPushingJWSTExtremes2025a} reference objects span $+0.5\sigma$ to $+4.3\sigma$ ($\mathrm{PTE} = 0.30$ to $1 \times 10^{-5}$), and the brightest, at $z = 16.3$, exceeds even the $f_{\star} = 1$ ceiling by $0.07$ dex.
Galaxy-to-galaxy burstiness raises this ceiling by a similar amount, $0.07$ dex (\app{burstiness}), so the object requires the conversion of essentially all of its halo's baryons into stars.

We do not read these offsets as evidence against $\Lambda$CDM, since the EVS prediction bounds the flux far more tightly than the photometry bounds the redshift.
A $+4\sigma$ offset is therefore most naturally resolved by an incorrect redshift determination; spectroscopic follow-up has indeed ruled out a number of extreme photometric candidates \citep[e.g.][]{zavalaDustyStarburstsMasquerading2023}.

\begin{figure*}
    \includegraphics[width=\textwidth]{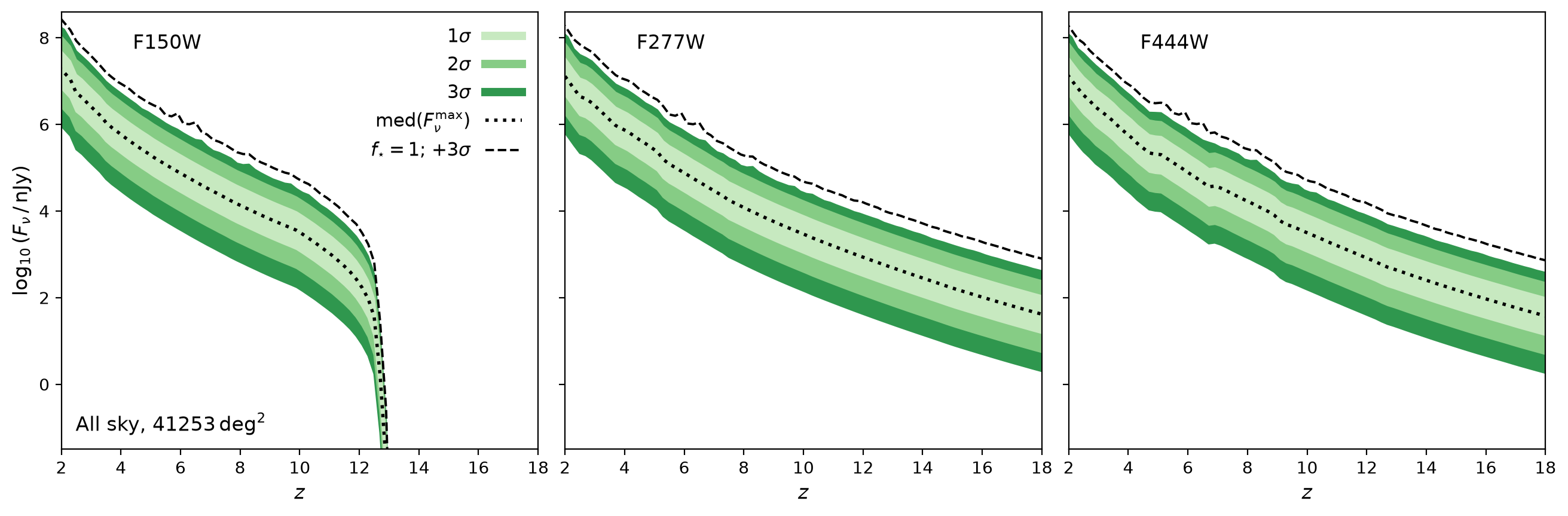}
    \caption{EVS flux predictions for the brightest galaxy in the whole sky ($f_{\mathrm{sky}} = 1$), in three JWST/NIRCam bands (columns). Shading, line styles and the fiducial model are as in \Fig{evs_confs}.}
    \label{fig:evs_fullsky}
\end{figure*}

\subsection{The brightest galaxy in the sky}
\label{sec:fullsky}

The comparison above is made against the projected volume of a particular survey.
It is also useful to ask what the formalism permits in our observable Universe: the flux of the single brightest galaxy on the whole sky in a given redshift shell.
We show this in \Fig{evs_fullsky}, computed identically to \Fig{evs_confs} but with $f_{\mathrm{sky}} = 1$.

The gain over the $0.28 \, \mathrm{deg^{2}}$ COSMOS-Web area is only $1.07$ dex at $z = 6$, rising slowly to $1.16$ dex at $z = 18$, despite the factor $1.5 \times 10^{5}$ in solid angle.
Equivalently, the median brightest galaxy in a COSMOS-Web-sized field lies only $\sim 0.2$ dex above the $-3\sigma$ contour of the all-sky distribution.
The practical consequence is that a candidate can only be shown to be genuinely anomalous by a large margin: an object exceeding the all-sky $+3\sigma$ contour would be inconsistent with the halo mass function irrespective of the area it was found in, but that contour sits $\sim 2.1$ dex above the median prediction for a typical JWST field, and no reported candidate approaches it.

\section{Discussion}
\label{sec:discussion}

The parametric SFHs of \sec{obs} are descriptive, and nothing in their definition prevents them from forming an unrealistically large fraction of a galaxy's stellar mass at late times.
The rapidly rising exponential, for example, forms most of its mass in the final e-folding time, which is why it gives the brightest predictions in \fig{evs_tau}.
Here we check whether the star formation rates (SFRs) implied by these SFHs could be supplied by gas accretion onto the host halo.

\begin{figure}
	\includegraphics[width=\columnwidth]{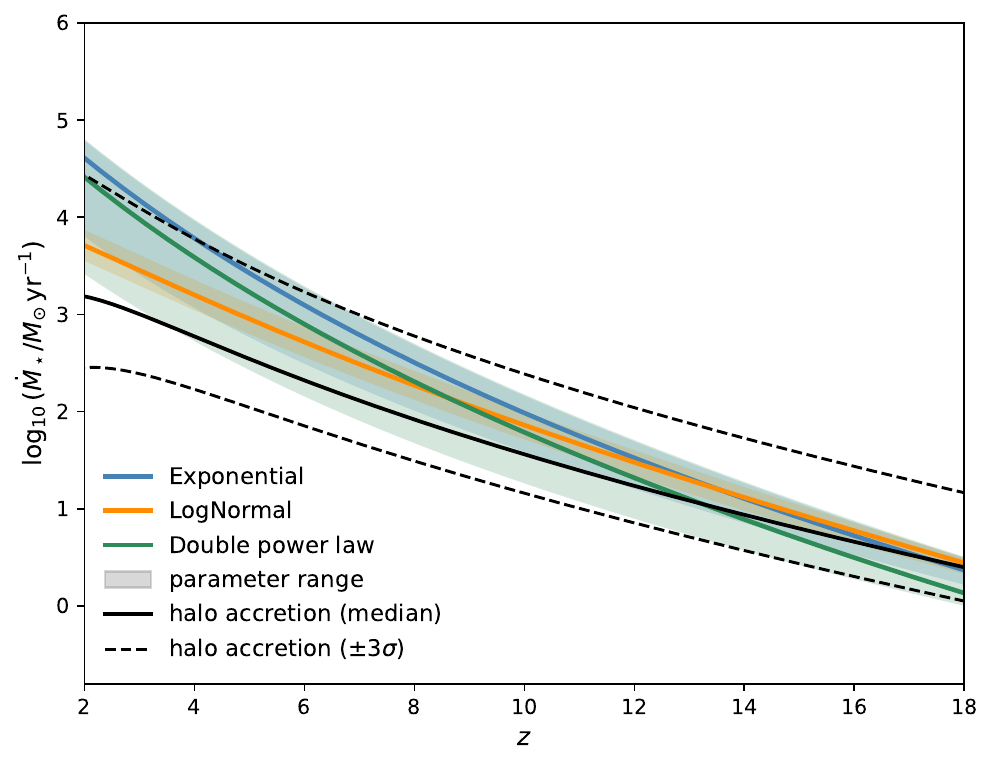}
    \caption{SFRs implied by the extreme tail of the EVS distribution, compared with the rate at which baryons are accreted onto the most massive halo in a $40\,\mathrm{arcmin^{2}}$ field. Black lines show $f_{\mathrm{b}}\,\dot{M}_{\mathrm{h}}$, using the median accretion rate of \citet{yungAreUltrahighredshiftGalaxies2024} for the median (solid) and $\pm 3\sigma$ (dashed) EVS halo mass. Coloured lines show the SFR of the $+3\sigma$ stellar mass for each SFH form of \Fig{sfh_models}, averaged over the preceding 100~Myr: fiducial parameters (solid) and the parameter range (shaded).}
    \label{fig:accretion}
\end{figure}

We compare two quantities in \fig{accretion}, both derived from the EVS distribution of the most massive halo in a $40\,\mathrm{arcmin^{2}}$ field.
The first is an upper limit on the SFR, set by the rate at which baryons are accreted onto that halo, $f_{\mathrm{b}}\,\dot{M}_{\mathrm{h}}$, where $\dot{M}_{\mathrm{h}}$ is the median halo mass accretion rate of \citet{yungAreUltrahighredshiftGalaxies2024}.
This limit assumes that every accreted baryon immediately forms stars, with no outflows and no gas retained in the interstellar medium.
It corresponds to an efficiency of unity, well above the $\lesssim 0.2$ typically inferred, and above the values proposed for feedback-free starbursts \citep{dekelEfficientFormationMassive2023,liFeedbackfreeStarburstsCosmic2024}.
The second is the SFR of the $+3\sigma$ stellar mass for each SFH form, averaged over the preceding $100\,\mathrm{Myr}$.

At $z > 6$ all three SFH forms lie below the accretion limit.
Towards higher redshift they converge on the median accretion rate, because the $100\,\mathrm{Myr}$ averaging window covers a growing fraction of the history ($40$ per cent of the age of the Universe at $z = 16$), so the averaged SFR depends less on the shape of the SFH.
The limit is exceeded only at lower redshift: by the fiducial exponential at $2.1 \lesssim z \lesssim 3.9$, and by the most rapidly rising exponential and double power law SFHs up to $z \simeq 6.5$ and $6.7$, respectively.
This is a limitation of these forms at late times, when an exponential with $\tau = 0.1\,\mathrm{Gyr}$ must form almost all of its mass in the last few hundred Myr of a multi-Gyr history.
The lognormal form, which has a finite width, stays below the limit at all redshifts.

The SFHs we use are therefore physically plausible over the redshift range of interest, and the spread in predicted flux found in \sec{systematics} does not rely on unphysical SFRs.
Likewise, the gap between the observed candidates and the upper contours in \fig{evs_confs} is not an artefact of the SFH choice.

\section{Conclusions}
\label{sec:conclusions}

We have extended the Extreme Value Statistics framework to predict the brightest galaxy in a given survey.
We couple the EVS halo PDF to a forward model of galaxy emission built on the \synthesizer{} framework to predict the full EVS distribution of the observed flux in a given photometric band.
This allows observed galaxy candidates to be evaluated directly in the flux--redshift plane, without relying on inverse modelling to recover stellar or halo masses for individual candidates.
Our main findings are:

\begin{itemize}
    \item We derive the extreme value distribution of the observed flux, by mapping the EVS halo mass distribution to stellar mass through a baryon fraction and a lognormal stellar fraction, and forward modelling the emission of the resulting population (\sec{evs}, \sec{obs}). Since the flux is linear in stellar mass, the confidence intervals of the flux distribution follow directly from those of the mass distribution.
    \item The choice of parametric star formation history is the dominant modelling systematic for the brightest-galaxy flux prediction, introducing a spread that grows with redshift and reaches $\sim 0.9\,\mathrm{dex}$ by $z \sim 18$ (\sec{systematics}). The double power law form, with a late peak time and shallow rising and falling slopes, gives the lowest $3\sigma$ flux limit, and is therefore the choice against which an observed candidate is most likely to appear in tension.
    \item Variations between widely used SPS models and the assumed IMF have little impact on the predicted flux distribution, contributing $\lesssim 0.15\,\mathrm{dex}$ and $\lesssim 0.20\,\mathrm{dex}$ respectively (\sec{systematics}). The treatment of nebular emission matters more than the choice of SPS model, raising the predicted fluxes by up to $\sim 0.26\,\mathrm{dex}$ for young, rising SFHs.
    \item The assumed stellar metallicity is a comparable systematic to the IMF: varying it over $10^{-4} \leq Z \leq 4 \times 10^{-2}$, up to the upper limit of the \textsc{bpass} grid, spans $\sim 0.3\,\mathrm{dex}$ (\sec{systematics}). All of the emission-model systematics combined therefore remain subdominant to the choice of SFH.
    \item Galaxy-to-galaxy burstiness allows a less massive halo to host the brightest galaxy.
    This raises the predicted flux by $\lesssim 0.1$ dex at both the median and the $+3\sigma$ contour, small compared with the SFH systematic, although the median shift grows with survey area to $0.26$ dex over the whole sky (\app{burstiness}).
    \item The luminous high-redshift candidates of \citet{caseyCOSMOSWebIntrinsicallyLuminous2024} cluster about the median prediction for the area from which they were drawn (\sec{lowz}). None is in tension with our fiducial model: every measurement lies at least $1.0$ dex below the $3\sigma$ upper contour. That margin is comparable to the SFH systematic, however, and under the most liberal double power law parameterisation the brightest COSMOS-Web candidate reaches the $3\sigma$ contour.
    \item Of the searches reaching beyond $z \sim 13$, the two spectroscopically confirmed galaxies at $z \simeq 14$ lie at $-1.2\sigma$ and $+0.1\sigma$ of the median for their area ($\mathrm{PTE} = 0.88$ and $0.48$), and are therefore entirely typical of what the halo mass function permits (\sec{highz}). The photometric candidates are far more extreme: the $z \sim 25$ candidates of \citet{perez-gonzalezRiseGalacticEmpire2025a} reach $+3.6$ to $+4.4\sigma$ ($\mathrm{PTE} = 2 \times 10^{-4}$ to $5 \times 10^{-6}$), and the brightest \citet{castellanoPushingJWSTExtremes2025a} object, at $z = 16.3$, lies at the $f_{\star} = 1$ ceiling, requiring the conversion of almost all halo baryons into stars. Since the prediction bounds the flux far more tightly than the photometry bounds the redshift, we regard such offsets as most naturally resolved by an incorrect redshift rather than as evidence against $\Lambda$CDM.
    \item Evaluated over the whole sky, the prediction sets an absolute ceiling on the brightest galaxy at a given redshift, independent of the survey considered (\sec{fullsky}).
    \item The star formation rates implied by the extreme tail of our parametric SFHs, averaged over $100\,\mathrm{Myr}$, lie below the rate at which baryons are accreted onto the host halo at $z > 6$ for every SFH form (\sec{discussion}). The limit is exceeded only at lower redshift, by the fiducial exponentially rising SFH at $2.1 \lesssim z \lesssim 3.9$, and by the most rapidly rising exponential and double power law SFHs up to $z \simeq 6.5$ and $6.7$, respectively; the lognormal form never exceeds it. The parametric forms used here therefore remain physically plausible over the redshift range of interest.
\end{itemize}

Forward modelling the EVS distribution provides a transparent treatment of the modelling assumptions that shape predictions of the most extreme galaxy candidates, and a complementary route to assessing the significance of extreme high-redshift candidates.


\section*{Acknowledgements}

The authors wish to thank Steven Murray for helping to incorporate the \citet{yungAreUltrahighredshiftGalaxies2024} mass functions in the \texttt{hmf} package.
We also thank Stephen Wilkins for helpful discussions.
CCL was supported by the research environment and infrastructure of the Handley Lab at the University of Cambridge.
The authors used Anthropic's Fable and Opus models to refine the code, and refine portions of the draft.
The authors take full responsibility for the final content.


\section*{Data Availability}

All data and codes used in the preceding analysis are provided at \url{https://github.com/christopherlovell/evstats}.




\bibliographystyle{mnras}
\bibliography{Extremes}



\appendix



\section{Sensitivity to the transfer function extrapolation}
\label{sec:hmf_transfer}

We compute halo mass functions with the \textsc{hmf} package \citep{murrayHMFcalcOnlineTool2013}, deriving the matter power spectrum from a transfer function computed with \textsc{camb} \citep{lewisEfficientComputationCosmic2000a}.
\textsc{camb} tabulates the transfer function only up to a finite maximum wavenumber, $k_{\mathrm{max}}$; beyond this it must be extrapolated to the small scales (high $k$) that set the abundance of low-mass haloes.
\textsc{hmf} provides two schemes for this: a linear extrapolation of the tabulated $\mathrm{log}\,T$--$\mathrm{log}\,k$ relation, or an analytic extrapolation using the \cite{eisensteinBaryonicFeaturesMatter1998} transfer function, matched to \textsc{camb} at $k_{\mathrm{max}}$.
The default was switched from the former to the latter for \textsc{hmf} versions later than 3.4.4 (the \texttt{extrapolate\_with\_eh} option), the analytic extrapolation being faster and, per the package developers, more accurate at low mass.

Although this choice nominally affects only the smallest scales, it has a surprisingly large impact on the extreme value prediction at high redshift.
In a small survey field the most massive halo at high redshift is itself of relatively low mass ($M_{\mathrm{max}} \sim 10^{9}\,\mathrm{M_{\odot}}$ at $z \sim 18$ for the areas considered here) corresponding to high $k$, precisely the regime where the two extrapolation schemes diverge.
The abundance of these extreme haloes, and hence the EVS prediction, therefore inherits a direct sensitivity to the transfer function extrapolation that grows towards high redshift: for our fiducial \cite{behrooziAverageStarFormation2013} mass function the effect is negligible ($\lesssim 0.02$ dex) at $z \sim 6$, but the newer prescription lowers $M_{\mathrm{max}}$ by $\sim 0.35$ dex by $z \sim 18$ (\fig{eh_comparison}).
As both schemes are extrapolations beyond the scales directly computed by \textsc{camb}, we regard this as a modelling uncertainty to bear in mind when interpreting the most extreme high-redshift masses.
We adopt the current default (\texttt{extrapolate\_with\_eh}\,$=$\,True) throughout.

\section{Halo mass function comparison}
\label{sec:hmf_comparison}

\begin{figure}
	\includegraphics[width=\columnwidth]{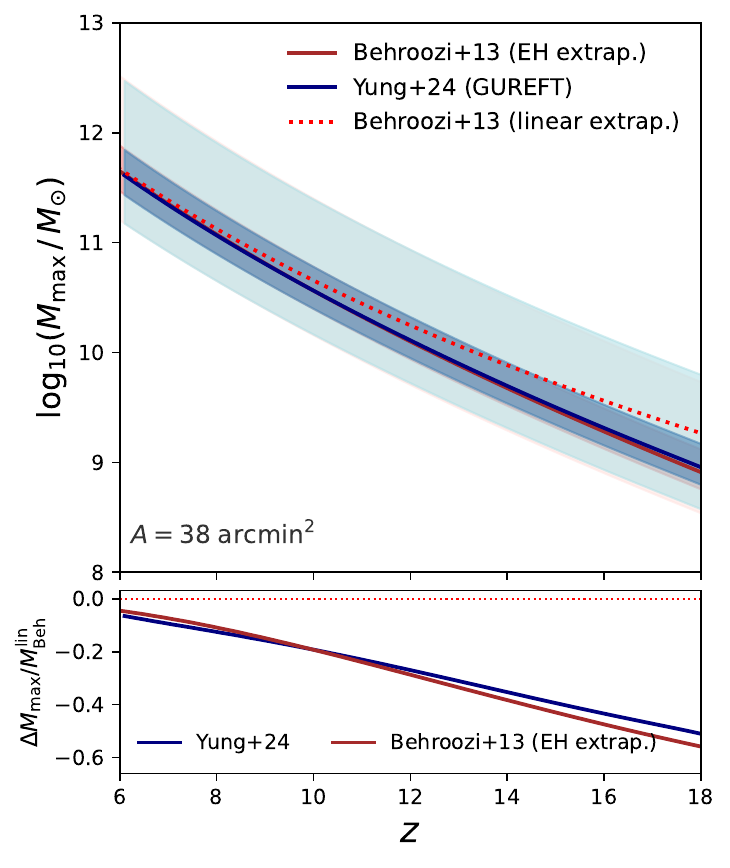}
    \caption{EVS prediction for the most massive halo mass in a $38 \, \mathrm{arcmin}^2$ field. We compare the EVS predictions given the \protect\cite{yungAreUltrahighredshiftGalaxies2024} GUREFT (blue bands) and \protect\cite{behrooziAverageStarFormation2013} (fiducial; brown bands) halo mass functions.
    Both are computed with the current \protect\cite{eisensteinBaryonicFeaturesMatter1998} transfer-function extrapolation.
    The dotted red line shows the median from \protect\cite{behrooziAverageStarFormation2013} assuming the previous linear extrapolation (the previous \textsc{hmf} default; see \app{hmf_transfer}). \textit{Bottom:} median fractional difference relative to the earlier prediction; the choice of fit contributes $\lesssim 10\%$, while the transfer-function update accounts for $\sim 55\%$ at $z \sim 18$.
    }
    \label{fig:eh_comparison}
\end{figure}

We adopt the \cite{behrooziAverageStarFormation2013} fit as our fiducial mass function, as used in \cite{lovellExtremeValueStatistics2023}, since it is defined across the full redshift range considered in this work.
Here we compare against the \cite{yungAreUltrahighredshiftGalaxies2024} GUREFT halo mass function, calibrated directly against $N$-body simulations over $z = 6$--$19$, rather than extrapolated to high redshift.
\fig{eh_comparison} compares the two: they agree to within $\sim 10\%$ across the calibrated range, so the choice of fit has little bearing on our predictions.

\section{Galaxy-to-galaxy variability in star formation history}
\label{sec:burstiness}

\Eq{flux_full} assumes that all haloes share the same stellar fraction and SFH, so that the most massive halo hosts the brightest galaxy.
To test this we superimpose stochastic fluctuations on each parametric SFH,
\begin{equation}
    \mathrm{log_{10}} \, \psi_{k}(t) = \mathrm{log_{10}} \, \psi(t) + \delta_{k}(t) \,,
    \label{eq:sfh_burst}
\end{equation}
where $\delta_{k}(t)$ is a damped random walk with amplitude $\sigma_{\delta}$ and correlation time $\tau_{\delta}$ \citep{caplarStochasticModellingStarformation2019}.
We adopt $\sigma_{\delta} = 0.24$ dex and $\tau_{\delta} = 50$ Myr, as found in the SERRA simulations \citep{pallottiniStochasticStarFormation2023}, and explore $0.12$--$0.5$ dex and $10$--$100$ Myr.
Realisations are normalised by the ensemble-mean mass, so that bursts change both the stellar mass and the light at fixed halo mass.
Without a limit, realisations above the mean would form more stars than their halo has baryons when $f_{\star}$ is close to unity, so we cap the stellar mass of each realisation at $f_{\mathrm{b}} M_{\mathrm{h}}$.
For the $f_{\star} = 1$ limit every realisation forms exactly $f_{\mathrm{b}} M_{\mathrm{h}}$, and bursts change only its light-to-mass ratio.

For $2 \times 10^{4}$ realisations per redshift shell we compute the offset in flux per unit mass from the smooth SFH, $\Delta = \mathrm{log_{10}} (\mathcal{F}^{X}_{\nu} / \mathcal{F}^{X}_{\nu, \mathrm{smooth}})$, by summing tabulated fluxes for each age bin of the \grid{} (\Eq{lnu}).
A galaxy with offset $\Delta$ is as bright as a smooth-SFH galaxy in a halo $\Delta$ dex more massive, so we replace the halo counts entering \Eq{cdf} with
\begin{equation}
    N_{X}(> \mathrm{log_{10}} M) = \left\langle N_{\mathrm{h}}(> \mathrm{log_{10}} M - \Delta) \right\rangle_{\Delta} \,,
    \label{eq:counts_burst}
\end{equation}
and then apply $f_{\star}$ and $\mathcal{F}^{X}_{\nu}$ as in \Eq{flux_full}, assuming that $\Delta$ is independent of halo mass over the narrow range of haloes that contribute.
Because the cap depends on $f_{\star}$, we evaluate \Eq{counts_burst} separately for each value of $f_{\star}$ before integrating over its distribution.

\begin{figure}
	\includegraphics[width=\columnwidth]{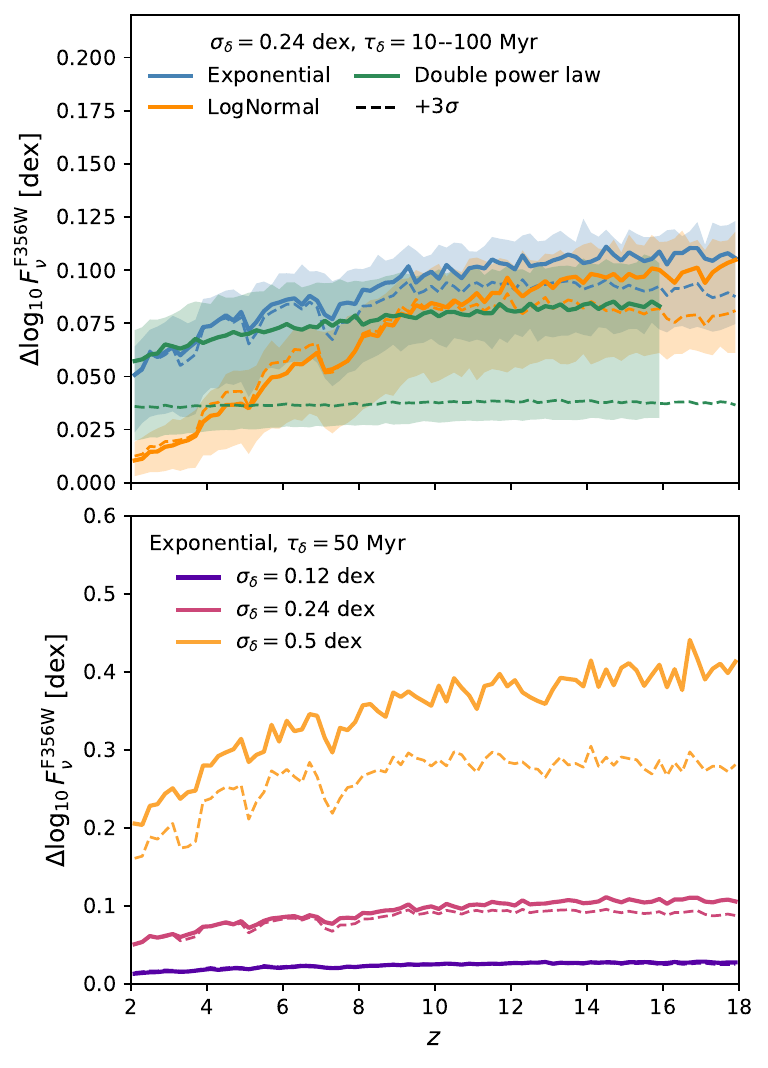}
    \caption{Shift in the median (solid) and $+3\sigma$ (dashed) EVS flux in F356W over the COSMOS-Web area due to galaxy-to-galaxy burstiness. \textit{Top:} each SFH form with $\sigma_{\delta} = 0.24$ dex; shaded bands span $\tau_{\delta} = 10$--$100$ Myr, and the double power law is truncated where its median flux falls below $1$ nJy. \textit{Bottom:} the exponentially rising SFH with $\tau_{\delta} = 50$ Myr. Small-scale structure is Monte Carlo noise.}
    \label{fig:burstiness}
\end{figure}

The resulting shifts are shown in \Fig{burstiness}.
For the fiducial burstiness the median and $+3\sigma$ flux increase by $\lesssim 0.11$ and $\lesssim 0.10$ dex for every SFH form and redshift, and the brightest galaxy is typically hosted by the second or third most massive halo.
The shifts are similar in other bands, since at these redshifts even the rest-frame optical light is dominated by young stars, and the bursts change the stellar mass as well as the light.
Stronger burstiness ($\sigma_{\delta} = 0.5$ dex) raises the prediction by $\sim 0.2$--$0.45$ dex, but implies a scatter in stellar mass at fixed halo mass of $0.16$--$0.35$ dex, at or above the $\sim 0.15$--$0.2$ dex typically inferred \citep{behrooziComprehensiveAnalysisUncertainties2010,enriquez-vargasEmpiricalEstimatesHow2026}.
The impact of burstiness is therefore small compared with the systematics of \sec{systematics}, although the median shift grows with survey area, to $0.26$ dex over the whole sky.
At $f_{\star} = 1$ bursts raise the ceiling by $0.07$ dex, similar to the amount by which the brightest \cite{castellanoPushingJWSTExtremes2025a} candidate exceeds the smooth-SFH ceiling.

\section{Extreme candidates in stellar mass}
\label{sec:stellar_mass}

\begin{figure*}
	\includegraphics[width=\textwidth]{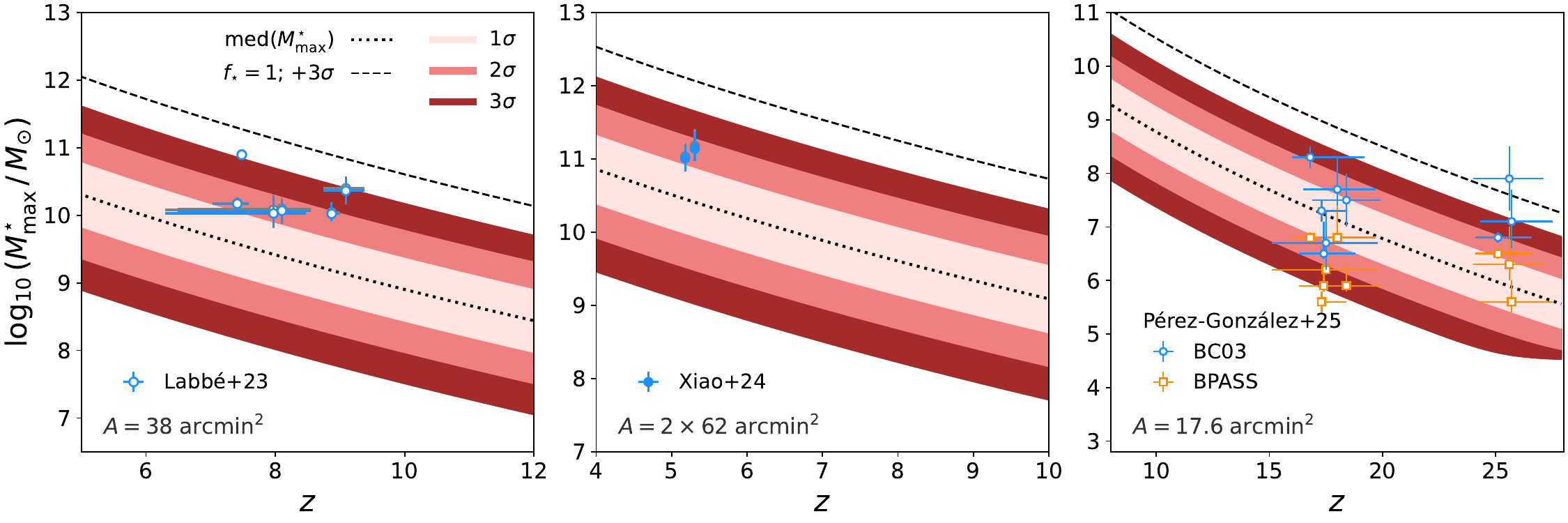}
    \caption{Extreme value predictions for the stellar mass of the most massive galaxy, compared to three samples of extreme candidates. Bands span the $1$, $2$ and $3\sigma$ EVS intervals, the dotted line the median and the dashed line the conservative $f_{\star} = 1$, $+3\sigma$ limit. Open and filled points denote photometric and spectroscopic redshifts, respectively. \textit{Left:} \protect\cite{labbePopulationRedCandidate2023} over $38 \, \mathrm{arcmin^{2}}$; grey points are the published masses, blue the same after the Eddington bias correction of \protect\sec{eddington}. \textit{Centre:} \protect\cite{xiaoAcceleratedFormationUltramassive2024} over $2 \times 62 \, \mathrm{arcmin^{2}}$. \textit{Right:} the $z \sim 17$ and $z \sim 25$ candidates of \protect\cite{perez-gonzalezRiseGalacticEmpire2025a} over $17.6 \, \mathrm{arcmin^{2}}$, with masses from the \textsc{bc03} templates adopted in that work (blue circles) and from \textsc{bpass} fitted to the same photometry (orange squares).
    }
    \label{fig:evs_mass}
\end{figure*}

We update the stellar mass comparisons of \citet{lovellExtremeValueStatistics2023} with the latest candidates (\Fig{evs_mass}), recomputed with the updated transfer function treatment (\app{hmf_transfer}).

The \cite{labbePopulationRedCandidate2023} candidates lie $0.6$--$1.4$ dex above the median. All but one lie within the $3\sigma$ interval; the exception, $\mathrm{log_{10}}(M_{\star} / \mathrm{M_{\odot}}) = 10.9$ at $z = 7.5$, sits at $+3.2\sigma$.
Their quoted mass uncertainties are small enough ($\lesssim 0.2$ dex) that the Eddington bias correction of \sec{eddington} moves them by at most $0.12\sigma$.
The ultra-massive dust-obscured galaxies of \cite{xiaoAcceleratedFormationUltramassive2024} are less extreme relative to their larger search area, at $+1.2$ and $+1.8\sigma$; we exclude their third candidate, S1, whose redshift was subsequently revised from $z = 5.58$ to $z = 3.2439$ once NIRSpec IFU spectroscopy showed the original grism measurement to be contaminated by a neighbour \citep{xiaoGrismIFURevising2026}.

These are, however, offsets in an \textit{inferred} quantity, whose systematic uncertainty typically exceeds the statistical uncertainty quoted alongside it.
The \cite{perez-gonzalezRiseGalacticEmpire2025a} sample makes this explicit: fitted to identical photometry, their \textsc{bpass} masses lie $\sim 1.5$ dex below their \textsc{BC03} masses, against quoted uncertainties of $0.2$--$0.6$ dex.
Adopting the former, all nine candidates lie between $-3.6$ and $+1.2\sigma$ and none is remarkable; adopting the latter they span $-1.6$ to $+4.6\sigma$, with the most extreme, at $z \simeq 25.6$, falling well outside the $3\sigma$ interval.
A candidate's position relative to the EVS distribution in stellar mass is therefore only as meaningful as the SPS model used to place it there; precisely what the flux comparison avoids.


\bsp	
\label{lastpage}
\end{document}